%% file: blood_supply_elasticity_Machado_Nov18.tex
\documentclass[11pt]{article}
 \pdfoutput=1
\usepackage[intlimits]{amsmath}
\usepackage{amsfonts,amssymb}
\usepackage{float,subfigure}
\usepackage{fancyhdr,fancybox}
\usepackage{framed}
\usepackage[toc,page]{appendix}

\makeatletter
\let\@xfloat@ORI\@xfloat
\makeatother

\usepackage{hyperref}

\makeatletter
\def\@xfloat#1[#2]{%
  \@xfloat@ORI{#1}[{#2}]%
  \def\baselinestretch{1}\@normalsize} 
\makeatother

\usepackage{geometry}
\usepackage{lscape}
\RequirePackage[longnamesfirst,authoryear,round]{natbib}
\usepackage{bm}
\usepackage{graphicx}
\usepackage{sectsty}
\renewcommand\bibname{References}
\usepackage{afterpage} 
\usepackage{verbatim}

\graphicspath{{Figures/}}

\usepackage{tabularx}
\usepackage{booktabs}
\usepackage{rotating}
\usepackage{eurosym}
\usepackage{tikz}
\usetikzlibrary{decorations.pathreplacing}
\usepackage[format=hang,labelfont={small,bf}]{caption}
\usepackage{authblk}

\usepackage{array}
\newcolumntype{L}[1]{>{\raggedright\let\newline\\\arraybackslash\hspace{0pt}}m{#1}}
\newcolumntype{C}[1]{>{\centering\let\newline\\\arraybackslash\hspace{0pt}}m{#1}}
\newcolumntype{R}[1]{>{\raggedleft\let\newline\\\arraybackslash\hspace{0pt}}m{#1}}

\usepackage{longtable}

\newcommand{\fn}{\footnote}

\newcommand{\tbf}{\textbf}

\newcommand{\beqn}{\begin{eqnarray}}
\newcommand{\eqn}{\end{eqnarray}}
\newcommand{\beqns}{\begin{eqnarray*}}
\newcommand{\eqns}{\end{eqnarray*}}

\newcommand{\bqt}{\begin{quote}}
\newcommand{\eqt}{\end{quote}}
\newcommand{\bitem}{\begin{itemize}}
\newcommand{\eitem}{\end{itemize}}

\DeclareSymbolFontAlphabet{\mathbb}{AMSb}
\setcaptionmargin{0.5in}

\newenvironment{mfignotes}{\begin{footnotesize}\begin{minipage}{\textwidth}\begin{scriptsize}\smallskip\par \emph{Notes -}  }
{\end{scriptsize}\end{minipage}\end{footnotesize}}
\newenvironment{mfigsource}{\begin{footnotesize}\begin{minipage}{\textwidth}\begin{scriptsize}\smallskip\par \emph{Source -}  }
{\end{scriptsize}\end{minipage}\end{footnotesize}}

\title{Estimating the Blood Supply Elasticity: Evidence from a Universal Scale Benefit Scheme}
\author[1]{Sara R. Machado\footnote{I thank my advisors Daniele Paserman, Johannes Schmieder, and Albert Ma. I thank Matteo Galizzi, Randy Ellis, Jawwad Noor, Daniel Wiesen, Matt Johnson, Kavan Kucko, Felipe Cordova, Osea Giuntella, Abigail Friedman, Pedro Pita Barros, Partha Deb, Marisa Miraldo, Eliana Barrenho, Richard Grieve, Mark Pennington, Bruce Hollingsworth, Luke Munford, Anindya Chakrabarti and S\o ren Kristensen for many valuable comments and conversations; Luis Negrao and Gracinda de Sousa from the Portuguese Blood Institute, for assistance in providing the data and for their availability; and participants at the BU Labor Reading Group, Health Economics Study Group (UK) Leeds and Lancaster, Portuguese Health Economics Association Conference, the Spanish Health Economics Association Conference, and seminar participants at BU, Harvard, LSE, LSHTM, Universidad de Navarra, and Imperial College for many helpful suggestions. 
I gratefully acknowledge  Ph.D. funding support from The Portuguese Science Foundation (SFRH/BD/68311/2010), as well as from the Institute for Economics Development at Boston University. }}

\affil[1]{  \footnotesize London School of Economics: Department of Health Policy; Old Building, Houghton Street, WC2A 2AE London (UK). Email: s.machado@lse.ac.uk }

\date{October 2018}

\begin{document}

\maketitle

\begin{abstract}

I estimate the semi-elasticity of blood donations with respect to a monetary benefit, namely the waiver of user fees when using the National Health Service, in Portugal. Using within-county variation over time in the value of the benefitI estimate both the unconditional elasticity, which captures overall response of the market, and the conditional elasticity, which holds constant the number of blood drives. This amounts to fixing a measure of the cost of donation to the blood donor. I instrument for the number of blood drives, which is endogenous, using a variable based on the number of weekend days and the proportion of blood drives on weekends. A one euro increase in the subsidy leads 1.8\% more donations per 10000 inhabitants, conditional on the number of blood drives. The unconditional effect is smaller. The benefit does not attract new donors, instead it fosters repeated donation. Furthermore, the discontinuation of the benefit lead to a predicted decrease in donations of around 18\%, on average. However, I show that blood drives have the potential to effectively substitute monetary incentives in solving market imbalances.

\end{abstract}

\clearpage

\setlength{\baselineskip}{4 ex}

\section{Introduction}
\label{blood_intro}

Blood is a prominent repugnant good. Similarly to kidneys, livers, and uterus, setting a price for blood is generally seen as inappropriate.\fn{\cite{RothRepugnance} defines repugnance as a particular form of distaste for certain kinds of transactions. The author points out its role as a market constraint, and provides a number of examples of such transactions. Slavery, usury, and body parts are examples of goods or services commonly treated as non-marketable.} For example, the law creating the Portuguese Blood Institute states 
	\begin{quote}
		{\em It is consecrated in this Law that blood will be provided for free from the moment of blood collection until blood transfusion to a patient in need. As a product of the human body, invaluable for both donors and other human beings, it should be banned from any form of market transaction.}
	\end{quote}
	
In practice, policymakers have explicitly refrained from giving monetary incentives.\fn{The history of the Market for Blood is well documented in \cite{MktBlood}. In developed countries, as well as most of the developing world, cash payments for blood are illegal due to ethical and safety concerns.} Market imbalances follow from the absence of a market-clearing mechanism. Blood shortages, defined as the supply of blood being below what is necessary for three days, are frequent.\citep{LaceteraAEJ12} Excess supply is not unusual, particularly in face of major events such as natural disasters, accidents or terrorist attacks.\fn{Blood has a 45-day shelf life. Moreover, storage has to follow strict safety guidelines, limiting available capacity. To make matters worse, a donor is suspended from donation at least for 90 or 120 days, respectively for men and women. A peak in the supply of blood can then be followed by a period of draught and supply shortages.}


Nonetheless, blood donors are sometimes given indirect incentives (e.g. badges, t-shirts, cups) to attract them to blood donation. \cite{ScienceBlood} and \cite{MktBlood} summarize the evidence, so far, regarding the effect of these extrinsic incentives: small stakes non monetary gifts increase blood donation.


In Portugal, a very important indirect incentive is the waiver of user fees for emergency rooms (ED) and other hospital and primary care services. Starting from 2003, the government strictly enforced the collection of user fees when using the National Health Service (NHS). However, regular blood donors are waived these user fees.\footnote{See Law 48/90 \citep{Lei48}, replaced by Law 27/2002 \citep{Lei27} and by Law 113/2011 \citep{Lei113}.} I study the effect of these benefit changes on the blood market. In other words, regular blood donors experienced a potential benefit from donation, and this benefit can be quantified. Furthermore, from 2012, emergency care no longer qualified for the user-fee waiver.


In this paper, I estimate blood supply elasticity with respect to the size of the user fee waiver. Learning about this parameter is key if we want to design policies that incentivize people to donate (either permanently, or in response to demand shocks). The benefit scheme under analysis has three particularly relevant features. First, it has been challenging to scale up to policy-making level or even run experiments with large samples of blood donors, due to difficulties in coordinating with blood collection services.\citep{Goette2010} In our study, we analyze the change in behavior of all potential donors assigned to the Lisbon Center of the PBI, which covers roughly half the country, resulting from a redefinition of the overall reward system for blood donations. Second, we are able to explore the role of an increase in the benefits for blood donations as well as its partial termination.\fn{The removal of incentives has been highlighted by \cite{gneezy2011} as one of the main challenges in studying how incentives impact prosocial behavior.} Third, and perhaps more importantly, the benefit is only active for regular blood donors. This serves a twofold purpose: one the one hand, to reward donors for their continued service to society\fn{According to informal sources at the Portuguese Blood Institute, as well as the Law Decree defining the benefit.}, and on the other hand, to foster regular blood donation.\fn{``An adequate and reliable supply of safe blood can be assured by a stable base of regular, voluntary, unpaid blood donors. These donors are also the safest group of donors as the prevalence of blood borne infections is lowest among this group. World Health Assembly resolution (WHA63.12) urges all Member States to develop national blood systems based on voluntary unpaid donation and work towards the goal of self-sufficiency''.\citep{WHOBlood}}


In fact, it's not even obvious that the elasticity is positive: some argue that in the context of blood donations and other charitable giving, extrinsic incentives crowd out intrinsic ones. For blood markets, this concern was raised in the seminal work by \cite{titmuss1971}. The author claims that, when agents are altruistic, providing rewards may backfire and decrease their willingness to participate in the market.\fn{See also \cite{BenabouTirole}, who identify image concerns as the main driver behind this effect.}  \cite{CrowdingBlood} revisit Titmuss' work, contributing to this debate with experimental evidence of the negative effect of cash transfers on women's willingness to donate. However, the results from \cite{LaceteraAEJ12, ScienceBlood,MktBlood} point in the opposite direction, with a positive impact of incentives on donations. The study by \cite{niessen2016} confirms these results, for direct cash payments in the south of Germany. Moreover, \cite{wildman2009} show that donors apparently do not exhibit pure altruism, based on a sample of Australian Red Cross blood donors.


I also study the elasticity of donations with respect to the number of blood drives. Blood drives are a different policy tool that blood collection agencies can use to generate more supply of blood. In practice, they give the organization more control over the available supply of blood.  They are similar to sales for blood donors, with lower transport cost, while providing a potentially higher social image benefit, when compared to donations at blood donor centers.\fn{Blood donor centers provide continuous demand opportunities at designated hospitals/Portuguese Blood Institute buildings. Regular blood drives take place with a set frequency at some onsite location (e.g., school, church, hall). There are also irregular blood drives. The first two provide a steady stream of blood supply. The third type can be thought of as peak load supply, typically activated when there is a blood shortage. They can also occur at other points in time, as a mechanism to attract new donors.}

In order to estimate blood supply elasticity, I use county-level variation over time in the value of  user fees and estimate the model as a difference-in-difference linear regression. 

The blood donations data come from a panel of the 241,605 potential blood donors enrolled in the Lisbon center of the PBI between 2003 and 2012. I analyze a total of 906,139 attempted blood donations. This dataset is unique in its ability to identify donors throughout the period, its large sample size, health data, and detailed information about blood drives.\fn{The dataset regarding user charges was built based on legislation, Health Care Delivery Reports and Reports and Accounts from a number of institutions within the National Health Service. These include the Central Administration of Health Care Services (ACSS), Sub-regional Health Administrations, Regional Health Administrations, Hospital Groups and Primary Care Center Groups, between 2003 and 2012.}

I estimate two types of blood supply elasticity. First, the unconditional elasticity, which tells us how the overall volume of blood donated to changes in the value of the user fee. It incorporates the response of the blood collection system to variation in the supply of blood. Second, the elasticity conditional on the number of blood drives, measuring the response of blood donors to the benefit, holding constant the cost of donation. 

However, the number of blood drives and blood donation are determined endogenously, downward biasing the estimates. I address this problem by instrumenting for blood drives. First, for each county, over the entire sample period, I calculate the proportion of drives that happened over weekends. Second, for each county and for each month, I calculate the total number of public holidays and weekend days. The instrument is the product of these two constructed variables, and is included in the differences-in-differences estimation. Furthermore, I estimate the elasticity of blood drives with respect to the user fee in a first-stage regression.

A 1\euro{} increase in user fees increases blood donations by 1.2\%, according to our estimates for unconditional blood supply elasticity. When conditioning on the number of blood drives, the estimate is 1.8\%. Moreover, the number of blood drives decreases as user fees increase. User fees and blood drives are substitutes in promoting donations. Changes in user fees affect returning donors most, as expected due to the design of the benefit scheme. User fees have a stronger effect on donations at blood drives than at donation centers. Finally, men respond more to the benefit, and young donors do not respond to the value of user fees.

The rest of the paper proceeds as follows. Section \ref{blood_background} describes the Portuguese blood market. Section  \ref{blood_descstat} presents the data and Section \ref{blood_metrics} the empirical strategy. Elasticity estimates and robustness checks are the subject of Section \ref{blood_results}.  Section \ref{blood_discussion} discusses the results and provides a cost benefit analysis of the user fee waiver. Section \ref{blood_conc} concludes.

\section{Institutional background}
\label{blood_background}

\subsection{Blood donation in Portugal}
\label{blood_bgd_donationPT}
The Portuguese market for blood donation is organized by a 100\% voluntary whole blood central planner, the Portuguese Blood and Transplantation Institute (PBI), a Government funded Agency within the Ministry of Health. The PBI is in charge of the collection, safety and delivery of blood to all health care institutions in the country. Since its creation, in 1990, any form of commercialization of blood is strictly forbidden, to the extent that hospitals are not charged for the blood they use when delivering health care.\citep{Lei48}
	
	Blood delivery and safety is a threefold challenge. Firstly, blood is an intermediate perishable good with a high depreciation rate, resulting in a 45-day shelf life. Secondly, the only way to produce blood is to extract it from living human beings who are willing to donate. Finally, demand and supply are largely unforeseeable, which gives rise to imbalances, both shortages and excess supply. 
	
	Therefore, a major task of the PBI is to make sure there are enough blood donation attempts so that enough blood is collected, but not so much at the same time that it will have to be destroyed. At the same time, the PBI has  efficiency concerns and aims to minimize the cost per unit of harvested blood.\fn{This has been one of the main objectives of PBI, according to their Annual Activity Report (namely 2010-2012, available at \url{http://www.ipst.pt/index.php/ipst-ip/instrumentos-gestao/pra}).} 
	
	 The PBI has traditionally resorted to two main instruments to minimize imbalances and recruit blood donors:  the organization of blood drives, and the waiver of user fees for visits to the National Health Service, which I define as the benefit.

\subsection{Blood drives}
\label{blood_bgd_drives}
Blood drives play a key role in preventing and/or solving market imbalances. In practice, blood drives provide incentives by decreasing the cost  and increasing the benefit of blood donation. In fact, due to social image concerns and other intrinsic motivation elements, they increase blood donors utility. Moreover, blood drives expose potential blood donors to the possibility of donating blood at a lower transport cost and higher visibility, which is believed to be important for prosocial behavior. The alternative to blood drives is to donate at a blood donor center, where there is much higher flexibility in the schedule, but also potentially higher transport cost and lower social visibility.  
	
	Blood drives can be organized directly by the Portuguese Blood and Transplantation Institute (PBI), or requested by either an individual or an association. Furthermore, the drives can take place at the PBI mobile blood collection truck or at some facility. These will typically be large buildings such as schools, office buildings, town halls, churches, etc.

\subsection{User fees for visits to the National Health System}\label{pt_ed}
\label{blood_bgd_userfees}
User fees, or user charges, are a type of healthcare copayment. Health care in Portugal is mainly provided by the tax-funded National Health System (NHS). Since the early 1990's, the Government legislates the values of user charges to be paid by every patient treated in the NHS, either inpatient or outpatient care. The copayments, capped at 1/3 of the total value of the procedure, are supposedly updated every year.

User fees, or user charges, are a type of healthcare copayment. Health care in Portugal is mainly provided by the tax-funded National Health System (NHS). Since the early 1990's, the Government legislates the values of user charges to be paid by every patient treated in the NHS, either inpatient or outpatient care. The copayments, capped at 1/3 of the total value of the procedure, can be updated every year. 
	
	User fees aim at preventing excessive demand for health care, while safeguarding access for those in need. They are particularly relevant for emergency care.\fn{See \cite{Barros12} for details on the role of user charges in the financing of the Portuguese NHS.} Due to equity concerns, since they were first defined, chronic patients, children, unemployed and other groups of population were waived these user charges. Regular blood donors and firefighters are the only non-underprivileged groups to receive this benefit.

	Until the Fall of 2003, user charges were a controversial issue, subject to lengthy political debate. At that moment, new legislation approved overruled all previous legislation regarding these copayments, establishing new rules and amounts to be paid.

	In April 2004, the Ministry of Health defined eligibility requirements to receive the waiver of user fees when visiting the NHS. Blood donors were required to have donated at least twice over the last 365 days. The blood donor card was defined as sufficient proof of eligibility.\fn{Before 2004, donors would have to request a document signed by the Portuguese Blood Institute representative.}  In January 2012, the waiver was restricted to non urgent care.\fn{Cfr. Law-Decree 113/2011 \citep{Lei113}} For blood donors this meant an increase in the copayment due for a visit to the ED. It should be noted that the policy change was motivated by a financial crisis. At the time, the government was forced to increase revenue and cut spending. User fees were one of the tools, together with a generalized tax increase, as part of a bailout plan.

\subsubsection{The Emergency Care Network Reform}
In 2008, the Ministry of Health announced the restructuring of the Emergency Care Network. As a result, each Emergency Departments (ED) was assigned one of three levels - Basic (SUB), Surgical (SUMC) and Multipurpose (SUP). Furthermore, as of February of 2009, the user fee was no longer computed based on the type of hospital/primary care center patients attended, but on the type of ED available at the location. 
	
The reform aimed to achieve better coordination and coverage in the provision of emergency care to the population. Since the National Health Service is centrally organized, i.e., there is no market defining where to place emergency care units, an emergency care network was designed, composed of the three levels of care defined above. The main optimization variable of the network, created by experts in health geography, was traveling time.
Within this network, any patient would be at a short driving distance from the most basic level of care, at least. 

To achieve the newly created network configuration, some hospitals saw their ED upgraded to provide a higher level of care and a set of primary care centers were integrated in the network as SUB. On the other hand, there was a set of hospitals and primary care centers with downgraded emergency services. The role out of these changes generates variation in the value of the benefit, within county, over time. For example, the upgrade of a Primary Care Center to SUB represents an increase in the {\em county user charge}, or the corresponding benefit for regular blood donors.

\section{Descriptive statistics}
\label{blood_descstat}
The data come from a panel of the 241,605 potential blood donors enrolled in the Lisbon center of the PBI between 2003 and 2012. I analyze a total of 906,139 attempted blood donations. The panel includes information regarding health (weight, height, blood pressure, hemoglobin, lab test results), the donation (site, staff, blood drive, waiting times) and socio-demographic variables (age, occupation, town).\fn{The dataset regarding user charges was built based on legislation, Health Care Delivery Reports and Reports and Accounts from a number of institutions within the National Health Service. These include the Central Administration of Health Care Services (ACSS), Sub-regional Health Administrations, Regional Health Administrations, Hospital Groups and Primary Care Center Groups, between 2003 and 2012.}

\subsection{Trends in blood donation and blood drives}
\label{blood_dstat_trends}
The number of attempted donations has increased since 2003, with a peak in 2009 at around 85000 donations. Of those, less than 75\% end up in approved blood collections, i.e., blood that can actually be used in the production of health care. (Figure \ref{donations_dc} a). Blood Donor Centers welcome around 20,000 potential donors each year. In 2012, this number decreased to levels similar to 2003. (Figure \ref{donations_dc} b) We observe a decreasing trend since 2008, which was further deepened in 2012, following the discontinuation of the subsidy. Also in Figure \ref{donations_dc}, we can see that the number of donations at blood drives grew until 2009. There was a decrease in the number of donations at blood drives in 2011, followed by a recovery in 2012.

Many potential donors are permanently denied donation, or at least postponed or suspended. Safety concerns regarding both donors' and patients' health are at the root of these decisions. To be able to donate blood, the donor has to be approved at a strict clinical triage process which includes a number of questions regarding lifestyle, risky behaviors, traveling, etc. Notice that in general this makes it harder to donate, particularly for younger people. The proportion of people who are suspended from donation is as high as 20\%. (Figure \ref{donations_app}) A concern from providing incentives for blood donation is the risk of an increase in the rejection rate. Although the number of suspended donors has not decreased, we don't observe any substantial increase in the number of eliminated donors, looking at raw data.

The data includes information about all drives organized by the Lisbon Center of the PBI. The area of influence is shown in Figure \ref{map}(a) , including the regions of Alentejo (Beja, Evora and Portalegre districts) and Algarve (Faro district), as well as the two districts surrounding Lisbon (the capital), Santarem and Setubal. Most blood drives occur in these three districts, the most densely populated ones. Figure \ref{map}(b) shows the distribution of the population by county, the unit of our analysis, at the 2011 Census.\fn{A district is a set of counties. Source: National Statistical Institute, 2012, available at \url{http://www.pordata.pt}.} In the analysis, I normalized donations or drives per count/month, dividing them by 10,000 inhabitants.

The number of blood drives varies greatly by district, as we can see from Figure \ref{drives_dist}. In 2012, there were more than 1500 blood drives in the Lisbon district, almost 5 per day, a large increase compared to the previous years. The number of drives stayed mostly constant over time in the other districts.

\subsection{User fees}
\label{blood_dstat_userfees}
User fees increased steadily between 2003 and 2012, in nominal terms.(Figure \ref{usercharges_ed}) The smaller adjustments reflected adjustments to inflation rate. The minimum amount paid per visit ranged from 2\euro{} in 2003 for a visit to a Primary Care Emergency Appointment to 20\euro{} due for a visit to a Multipurpose Emergency Department in 2012. The copayments doubled in 2012. For example, a primary care emergency appointment increased from 4.5\euro{} to 10\euro{}, in January 1, 2012. 
Regular blood donors face an even higher price increase, since previously they didn't have to pay. 

As we can see in the picture, until February 2009, the user fee was determined by the type of health care facility (Central Hospital, District Hospital, District Level One Hospital or Primary Care Center). From then on, it was based on the type of Emergency Department available at the healthcare unit. This change was a consequence of a more fundamental policy reform and generated important variation in the value of the benefit for regular blood donors.

\subsection{Cross-district variation in blood donations and blood drives}
\label{blood_dstat_counties}

In order to measure the effect of the user fees waiver, I analyze within county variation in the value of these fees. Portugal is divided into 18 districts, which further divide into counties. The organization of the provision of health care services within the Ministry of Health follows these geographical divisions. For example, there is at least one Primary Care Center per county and in many cases there will be exactly one. Furthermore, networks of care and referral patterns are organized within districts. Therefore, the Emergency Care Network also takes into account these geographical groups. Since I am able to identify in the data each donor's county of residence, that allows me to uniquely identify each donor's local user fee.

We follow all counties (the cross sectional component) over time (a monthly time series) and we measure the number of donations made by people who live in that county, as well as the number of blood drives hosted in that county. 

The number of blood donations by county/month, per 10,000 inhabitants, our dependent variable, increased between 2003 and 2010. In Figure \ref{avg_pcap_don_dist} we see the evolution of the average of this variable, between 2003 and 2012, by district. Lisboa and Santarem have the highest average of donations per 10,000 inhabitants, per month, between 20 and 25 donations over most of our sample. Beja and Setubal have between 10 and 15 monthly donations per 10,000 inhabitants. The remaining districts have the lowest donation rates, below 10 and even 5 donations per 10,000 inhabitants. Furthermore, the distribution of the number of blood donations became more spread out over time ($\sigma_{2003} = 17.43$, $\sigma_{2012} = 21.14$), but there is a core group of counties that remain in the upper quartile of the distribution throughout the whole period, mostly in the districts of Lisboa and Santarem.  

When we look at the average number of blood drives per month, per 10,000 inhabitants, we find a similar geographical pattern, with one exception.(\ref{avg_pcap_drives_dist}) Beja, a district with a relatively high number of donations per 10,000 inhabitants, has fewer drives, compared to the other districts. There were 1.2 drives ($\sigma=4.50$) and 57.43 ($\sigma=129.53$) donations per county/month, on average.\fn{We measure the number of drives per county as the number of blood drives that happened in county $i$ in month $t$. Notice that blood donations per county/month are donations made by residents of county $i$, not necessarily in the same county.} 
 The first quartile of distribution does not host blood drives. Only in the 4th quartile of the distribution do we find counties with an average of more than one blood drive per month. There is variation across counties and within counties in the distribution of blood drives. The mean decreased in 2012, to 0.18 drives per 10,000 inhabitants.

\subsection{User fees and blood donations}
\label{blood_dstat_userfees2}

Furthermore, we measure the maximum possible value of the user fee to be paid in that county, 

The main explanatory variable, maximum user fee, is equal to the highest baseline user fee in county $i$ in month $t$, which depends on the type of Emergency Department available.\fn{Notice that the value of the user fee that I am using is the one a patient has to pay for the visit. Additional fees may be charged for diagnostic and treatment procedures. The total payment is caped at 50\euro. In any case, beforehand, the patient only knows with certainty that she has to pay this baseline user fee. This can be thought of as buying insurance. The minimum value saved is this user fee, which varies across healthcare units, and the maximum is 50\euro. To build this variable I used a number of different sources, which include the Statistics of Health Care Provided of the Ministry of Health, Annual Reports and Accounts of Regional and Local Health Services.}  

The value of the user fee is defined as the ED benefit available in the county. The amount due first depended on the type of Hospital or Primary Care Center being visited. From February 2009 on, it was based on the type of Emergency Department the patient went to. In the cross-sectional dimension, we measure what type of Emergency care is available in each county and compute the corresponding user fee. We then do the same for all months in our analysis. The opening or closure of Emergency Departments give rise to time-series variation in the user fee. Given the nature of the policy change, changes in the Emergency Care Network are exogenous to the number of blood donations in the county. 

In Figure \ref{avg_donations} we see that the direction of the relationship between the average number of donations and the value of the user charge is quite unclear. The number of Blood Drives shows a negative relationship with the benefit.(Figure \ref{avg_drives}) When looking at specific counties, we sometimes observe an increase in the number of donations coincides with the increase in the value of the user charge. It also seems to be the case that, in some counties, there is a sharp drop in donations when the waiver is removed. This drop is steeper in the other centers of the PBI (Porto and Coimbra), not included in this study. In Lisbon, in contrast, the removal of the benefit coincides with an increase in the number of blood drives. That is to say, there was a transfer of blood drives from the periphery to the capital.

\section{Empirical Strategy}
\label{blood_metrics}


The model to estimate the unconditional blood supply elasticity is given by
\begin{equation}\label{eq_unconditional}
	\text{donations}_{i,t}= \beta_0 + \beta_1 \text{benefit}_{i,t} + \gamma_i + \delta_t +\epsilon_{i,t}
\end{equation}
where $i$ is the county and $t$ is the month. The number of donations is normalized by the population of the county (times 10000). 

The coefficient $\beta_1$ measures blood supply elasticity, taking into account the response by the Portuguese Blood Institute to shifts in supply and in the organization of blood drives, using within-county variation over time. This coefficients tells us the overall effect of the value of the user fee on blood donations.


We are also interested in estimating the conditional blood supply elasticity, given by
\begin{equation}\label{eq_conditional}
	\text{donations}_{i,t}= \beta_0 + \beta_1 \text{benefit}_{i,t} + \beta_2 \text{drives}_{i,t}+ \gamma_i + \delta_t +\epsilon_{i,t}
\end{equation}
where $i$ is the county and $t$ is the month.

In this case, $\beta_1$ measures blood supply elasticity, controlling for the number of blood drives held in county $i$ at month $t$. That is to say, holding constant the cost of donation, it measures the supply response to changes in the value of user fees. This allows us to isolate the response of individual blood donors to the incentive, and separate it from the market's response.

However, this raises a different problem. Clearly, the number of blood drives is endogenous. There will be more blood drives in places where the Portuguese Blood Institute expects more people to show up to donate. The estimate for $\beta_1$ will be downward biased, given the negative relationship between blood drives and user charges (see Figure \ref{avg_drives} and Section \ref{blood_results_conditional}).

I use an instrumental variable based on whether a blood drive is held on a weekend or on a weekday, one of the main characteristics of blood drives. The instrument is defined as 
\beqn
	IV:Weekend_{it}= \text{\# Weekend Days}_{it}  \times \frac{1}{T} \sum_{t=1}^T \frac{\text{\# weekend blood drives}_{it}}{\text{\# blood drives}_{it}},
\eqn
where $i$ is the county, $t$ is the month, and $T=120$, the total number of months in our sample. The two components of the instrument are the number of weekend or holiday days in a given county/month, and the average proportion of drives held on weekends, for county $i$.

The number of weekend days in a county/month vary, on average, between 8.5 and 11.5 days (Figure \ref{avg_weekendays}), across counties, between 2003 and 2012. To build this variable, in addition to weekends, I included local county holidays, as long as they didn't fall on a weekend. 

Focusing on the second component of the instrument, we can see that there are more drives on weekends than on weekdays.(Figure \ref{drives_time}) 
In Figure \ref{drives_type} we can see that there is a systematic relationship between weekend drives and the type of blood drive location. The same holds true for weekend drives and geographical location, as shown, for example, in Figure \ref{drives_type_d3}, for the counties in Lisbon's district. One clear picture emerges from these figures and that is the relevance of the type of day of the week in the number of blood drives. 

When estimating the conditional blood supply elasticity, we will use the variation in $IV:Weekends$ to identify the effect of the number of blood drives on blood donations, {\em conditional} on county and month fixed effects, which absorb the variation due to county specific characteristics. 

The main concern regarding the validity of the instrument would be the exclusion restriction. This condition requires the number of blood donations per 10,000 inhabitants to be uncorrelated with {\em IV:Weekend}. 

First, notice two reasons as to why the decision to hold blood drives, even if correlated with the number of potential blood donations, differs from the decision to donate. The first has to do with regularity of donations. A donor faces medical constraints that determine when he can donate. For instance, there is a minimum interval between donations, three months for men and 4 months for women. There are also clinical reasons behind the rejection or deferral of a potential donor for a period of time, such as traveling to a tropical country or due to blood test results. The second reason is the how the central planner organizes blood drives, which are planned ahead to try to minimize imbalances and the cost per unit of blood harvested. The higher the demand for blood, the higher the number of blood drives, but as much as possible as a result of careful forecasting of stocks and planning ahead. Some of these blood drives will occur on weekends, and some on weekdays.
As for a donor's decision, even if he has a higher propensity to donate on weekends, the decision is likely to be made between donating today, Wednesday, or over the weekend, taking into account blood donor center and blood drives availability and costs. 

In practice, the threat to identification comes from counties that have a particular characteristic that makes people donate more/less on weekends, and that determines whether there is a blood drive there on a weekend or not. It is  possible that the number of weekend days in itself be somewhat correlated with the number of donations per 10,000 inhabitants. Given that we are using within-county variation in the estimation, this problem does not undermine the results.

\section{Results}
\label{blood_results}

Throughout the analysis I estimate blood supply response to user fees using two sets of models, one in levels and one calculating the semi-elasticity of donations with respect to the user fees. The dependent variable is highly skewed to the left.(Figure \ref{hist_don}) To estimate the semi-elasticity, I have transformed the dependent variable, donations per 10,000 inhabitants, using the inverse hyperbolic sine transformation ({\em arsinh}, or IHS)\fn{One advantage of the IHS is its domain, the real line.}:
\beqns
	\text{IHS donations}_{it}= \ln \left( \text{donations}_{it} + \sqrt{\text{donations}^2+1} \right).
\eqns

Therefore, the coefficient $\beta_1$ is the average change in blood donations resulting from a one euro increase in the benefit, measured in either donations per 10,000 inhabitants (levels) or as the growth rate (IHS).

\subsection{Unconditional elasticity}
\label{blood_results_unconditional}

The estimates for Eq \ref{eq_unconditional} are presented in Table \ref{tab_all}. The left hand side shows the results in levels.  A one euro increase in the benefit increases donations by 0.12 donations per 10,000 (Column (2)), when controlling for both county and month fixed effects. This is about 1\% of the mean of the dependent variable, however, it is imprecisely estimated. The point estimate is slightly lower, 0.097, when we control for district-specific trends (Column (3)). 

The estimates on the right-hand side can be interpreted as the percent increase in donations resulting from a one euro increase in the benefit. Donations increase by 1.2\% (Column (2)), conditional on county and month fixed effects. The effect is higher when we control for district-specific trends. The remaining columns test for different specifications for county/month fixed effects.\fn{I estimated a series of models using alternative specifications, for both the unconditional and the conditional model. Log-transforming the dependent variable is sensitive to the value of the constant added to avoid $\log(0)$. I have also estimated a two-part model, to see whether there were two separate processes generating the data, one for the zeroes, i.e., whether or not to participate in the market, and one for the actual choice of quantity. The results go in line with what we estimated previously, but they don't fit the data particularly well. Finally, the number of blood donations could be thought of as an example of count data. The estimates from poisson and negative binomial estimation do not seem to describe the actual data generating process. Results are available upon request.}

\subsection{Conditional elasticity}
\label{blood_results_conditional}

We now turn to the estimation of Eq. \ref{eq_conditional}, using both OLS and 2SLS. First, we will study the impact of the instrumental variable and of the user fee on the number of blood drives, the endogenous variable. 

The first stage regression is given by
\begin{equation}
	\text{drives}_{i,t}= \beta_0 + \beta_1 \text{benefit}_{i,t}  + \beta_2 \text{IV:Weekend}_{i,t}+ \gamma_i + \delta_t +\epsilon_{i,t}
\end{equation}
where $i$ is the county and $t$ is the month and the number of drives is normalized (per 10,000 inhabitants).

The estimation results are shown in Table \ref{tab_fstage}. The number of blood drives decreases with the benefit and increases with the instrument. A one standard-deviation increase in the instrument generates 25.8\% more blood drives. On the other hand, a one standard deviation increase in ED benefit leads to a 2\% drop in the number of blood drives, on average. These results are based on Column (2), which uses the instrument in levels. The instrument has a large mass at zero, due to counties that have zero weekend drives. I estimate Column (3) to take into account this nonlinearity in the regressor. 

The results from the second stage estimation of Equation (\ref{eq_conditional}) are shown in Table \ref{tab_sstage}. All specifications include county and calendar month fixed effects. The point estimate for the effect of the benefit on blood drives more than doubles once we condition on blood drives (Columns (1)-(2)). The 2SLS estimate is even higher (Columns (3)-(4)). The results are similar for both specifications of the instrument. When the user fee increases one euro, we estimate a 1.8\% increase in donations per 10,000 inhabitants, on average (Column (3), right hand side). The model is well identified (Weak Identification F-statistic of 207.4).

\subsection{Robustness checks and extensions}
\label{blood_results_robustness}

The set of results shown in Tables \ref{tab_checks} and \ref{tab_sstage_checks} aims at analyzing whether the results are driven by a particular set of observations, for unconditional and conditional elasticity, respectively. 

We first estimate blood supply elasticity restricting the sample to the years before 2012, before the user fee waiver was removed. The point estimate for the unconditional semi-elasticity of donations per 10,000 inhabitants with respect to the benefit is practically the same, a 1.1\% growth in donations. However, excluding 2012 has an impact in the estimate. The semi-elasticity with respect to the benefit is much smaller when we condition on blood drives. 

Lisbon, the capital's district, has more donations per capita than the other districts. It also has a higher density of blood drives and blood donor centers. Excluding donations in Lisbon yields a slightly higher elasticity than the one found before, around 1.7\% (Table \ref{tab_checks}). Counties in the periphery respond more to changes in the benefit, in the unconditional model. However, the point estimate for is exactly the same if we exclude Lisbon from the analysis when estimating the conditional elasticity (Table \ref{tab_sstage_checks}, Column (1)-(2)).

\subsubsection{Blood supply response mechanism}
In Table \ref{tab_samples} we find unconditional semi-elasticity estimates for a set of donor subgroups. Looking at different donor subgroups gives us a deeper insight into the mechanism driving blood supply's response to change in the user fee. All models include county and month fixed effects. In Table \ref{tab_sstage_samples} we replicate the analysis done in Table \ref{tab_samples}, this time for the conditional elasticity estimates.

\paragraph{Extensive versus intensive margin}
Providing small stakes incentives should increase blood donation.\citep{ScienceBlood, MktBlood} We now ask the question as to whether incentives affect the extensive margin, i.e., are new donors attracted to donation because there is a subsidy involved? 
According to these results (0.0026, Column (2) of Table \ref{tab_samples}), the effect of the subsidy is much smaller for new donors, which would indicate that most of the effect is in fact on the intensive margin, and not in the extensive. 
Notice that the incentive scheme is designed to benefit regular blood donors. These estimates are telling us that those donors who were already willing to donate, donate more because of the benefit. In other words, new donors are not attracted to donation because of the benefit, but once they are donors, they might {\em stay more} because of the user fee waiver. This is confirmed by the conditional elasticity estimates. The benefit has a small and imprecisely estimated impact on the number of new donors (0.4\%, Table \ref{tab_sstage_samples}, Column (3)).

\paragraph{Socio-Demographic Differences} 
We further investigate whether men and women respond similarly to the benefit. \cite{CrowdingBlood} find evidence of higher crowding out for women. In this case, donations by men are significantly more responsive to the benefit (Columns (4)-(5)). Finally, I analyze the semi-elasticity of different age groups. There could be differences in demand for health care services between younger and older age groups. I find that the younger donor group (Column (1) of Panel B, Age$<25$) do not respond to the benefit. The remaining age groups are statistically equivalent and have the same semi-elasticity as the full sample. 

Men respond more to the benefit, also when we condition on the cost of donation (Table \ref{tab_sstage_samples} Columns (4)-(5)). A one euro increase in the subsidy leads to 0.31 more donations per 10,000 inhabitants, at the mean of the dependent variable. Once again we verify that the youngest age group does not react to the incentive, even conditional on blood drives, and the other age groups are statistically equivalent.

\section{Discussion}
\label{blood_discussion}

We have studied blood supply elasticity with respect to changes  in the value of the user fee. At this point, we focus on a particular aspect, the removal of the waiver for ED user fees. The estimates change significantly when we restrict the sample to the years before 2012, at which point the benefit was removed. In Table \ref{tab_removal} we study the semi-elasticity estimates when we restrict the sample to 2011 and 2012. At this point, most of the changes in the benefit were its removal in the beginning of 2012. The unconditional estimates are very similar to the ones in Table \ref{tab_all}. However, two-stage least squares estimates yield a higher semi-elasticity, 2.1\%, on average (Column (2) and (4)). 

Therefore, the behavioral response of donors was stronger when the waiver was removed. The blood market compensated for this change by decreasing the cost of donation to donors, increasing the number of blood drives, which leads to an overall lower unconditional elasticity of blood supply. 

From a policymaker point of view, we would like to now how the removal of the waiver would that translate into additional revenue for the National Health System. 

According to the Ministry of Health, on average a person will go to the Emergency Department every two years. We can take the number of donors with blood collections in 2011 as the potential number of regular blood donors, eligible for the benefit in 2012. That amounts to  35000 donors. Of those, half would go to the hospital in 2012. On average, the (baseline) user fee would be 10\euro. Therefore, the NHS would be able to raise a minimum of 175,000\euro, on average, by removing the ED waiver. 

Moreover, the removal of the benefit has another advantage for ED. Around 40\% of ED visits are marked as green or below, measured by the Manchester Triage System. In practice, this means that 40\% of the people that visit the ED should be going to a less urgent level of care. This  creates congestion problems in ED. If the user does not have to pay the user fee, the likelihood of making an unnecessary ED visit is even higher.

These are the benefits of removing the user fee waiver. On the cost side, we have a potential increase in the cost per blood unit. This happens because the central planner has to compensate for the decrease in the number of blood donations with an increase in blood drives. This has a direct impact in labor costs, since the demand for health care professional staff (nurses, physicians and lab technicians) increases. The Portuguese Blood Institute has a set number of staff on payroll, and resorts to per diem professionals when the PBI's staff is not enough to staff the necessary blood drives. Therefore, a higher number of blood drives increases the number of per diem workers. Furthermore, many of these additional drives will happen on weekends and after work hours, which means they are paid at a higher  wage rate.  

The PBI reports large increases in staff overtime remuneration, in its 2012 Annual Activity Report. Namely, there was a 29.03\% increase in overtime payroll expenditure. In order to staff a blood drive, at least a physician and a nurse are needed. The 400\% increase in physician's overtime compensation (25.89\% increase for nurse's), from 38,666\euro{} to 193,817\euro{}, reflects the increase in the cost per unit of blood. 

In practice, more blood drives also implies less efficient blood drives. By less efficient the Portuguese Blood Institute means drives with a higher cost per unit of blood harvested. This could be due to the increase in cost, and to a lower number of donors per drive. In 2011, the is unit cost os 172.73\euro. Unfortunately this data for 2012 is not available, so we can't confirm that there was an increase in costs and decrease in efficiency. 

It is therefore unclear whether the removal of the waiver was cost effective or not. However, it is clear that changing the benefit scheme for blood donors had an impact in the blood donation market that goes beyond individuals' decision of becoming blood donors.

\section{Conclusion}
\label{blood_conc}
We analyzed the impact of a benefit used by the Portuguese Blood and Transplantation Institute to foster blood donations. We estimated the semi-elasticity of blood donations with respect to user fees. To do so, we used changes in the Emergency Care Network to establish causality using a differences-in-differences approach. Furthermore, we estimated both the unconditional elasticity, which captures overall response of the market, and the conditional elasticity, which holds constant the number of blood drives. This amounts to fixing a measure of the cost of donation to the blood donor. We dealt with the problem raised by  the endogeneity of the number of blood drives, using an instrument based on the number of weekend days and the proportion of blood drives on weekends.

We find that the benefit for regular blood donors increases the number of blood donations. A one euro increase in the subsidy leads 1.8\% more donations per 10000 inhabitants, conditional on the number of blood drives. The unconditional effect is smaller. Younger donors do not react to the waiver of user fees. The benefit does not attract new donors, instead it fosters repeated donation.

The discontinuation of the benefit lead to a predicted decrease in donations of around 18\%, on average, since the average value of the user fee in 2012 was 10\euro{}.  This had to be compensated by a large increase in the number of blood drives, with an increase in the cost per harvested unit. 
 
On the other hand, this change in the incentives could lead to savings of at least 175,000\euro{} per year for the National Health Service. It is feasible to think of a scenario in which these savings lead to a bigger investment in blood drives., or in the optimization of the system to minimize imbalances. Given the results in this study, blood drives could effectively substitute for the decrease in donations resulting from the discontinuation of monetary incentives, particularly as a mechanism to solve market imbalances. It might not be worth it, in the end, to subsidize blood donations, particularly if the subsidy ends up being temporary. However, the efficiency cost of additional blood drives has to be taken into account. 

Nonetheless, further analysis has to be done. We have treated all agents as neoclassical, without any assumptions regarding behavioral reasons driving behavior. This paper deals with the supply and demand of blood donations. Counties which have higher net benefits of donations are shown to have a higher volume of donations. In future work, we will explore individual-level data to answer the question of how incentives for repeated donation affect the behavior of donors. Specifically, we are interested in analyzing which donors respond to deadlines in the incentive and how that affects their donation behavior.



\input{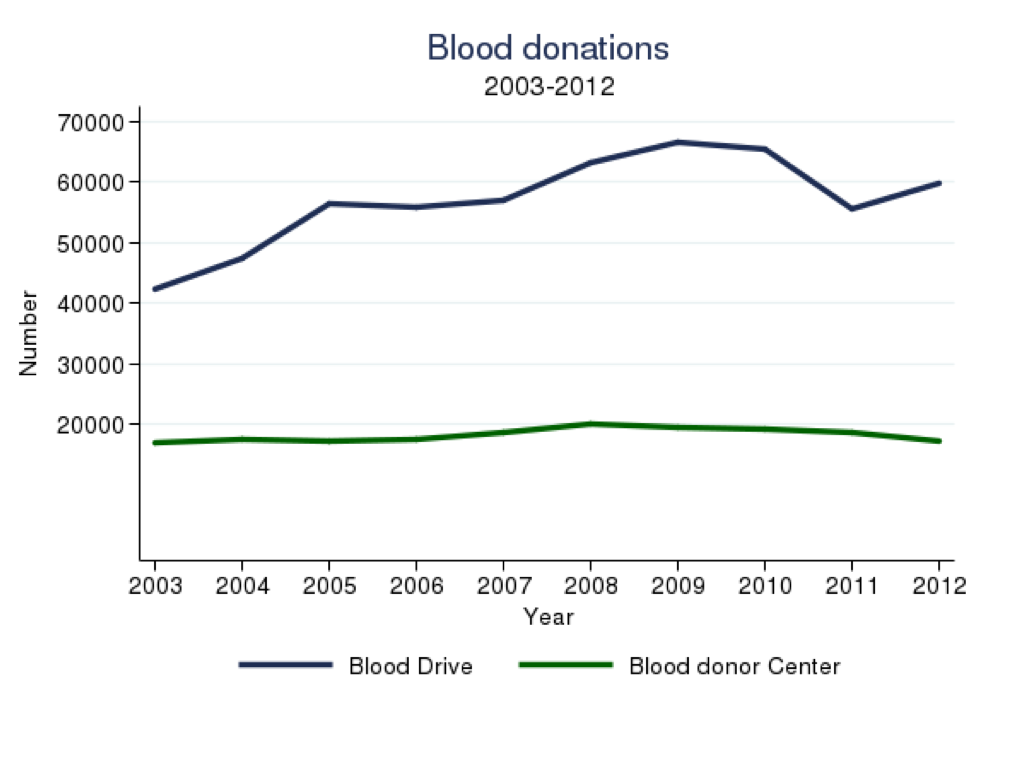}

\input{donations_app}

\input{map}

\input{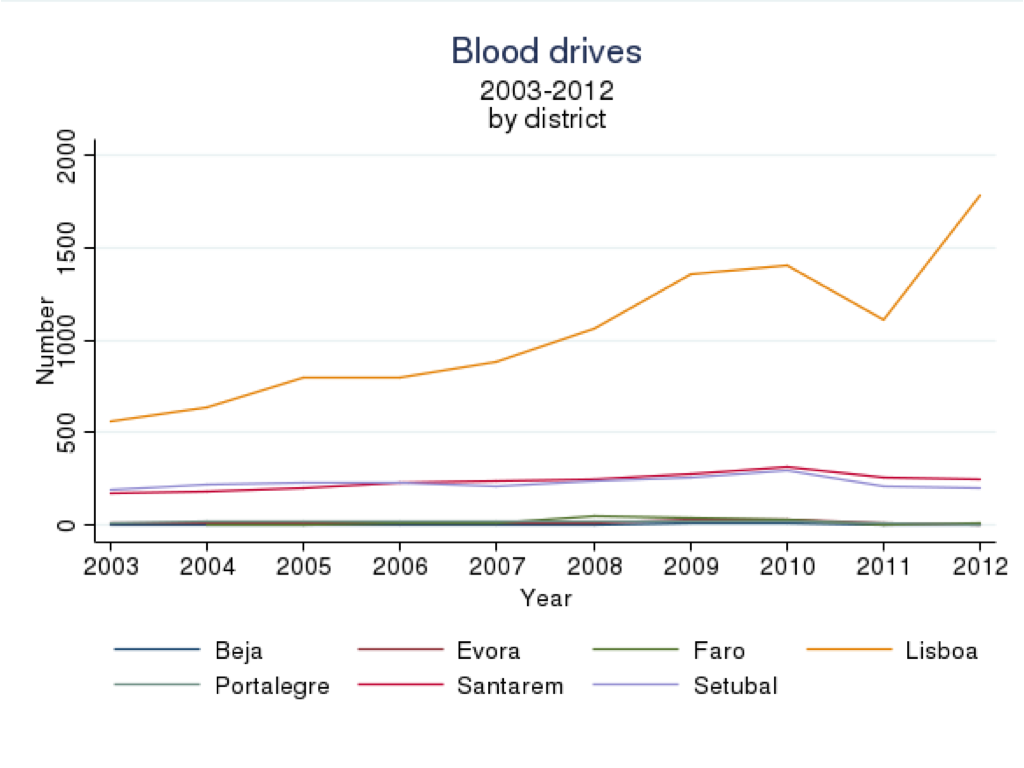}

\input{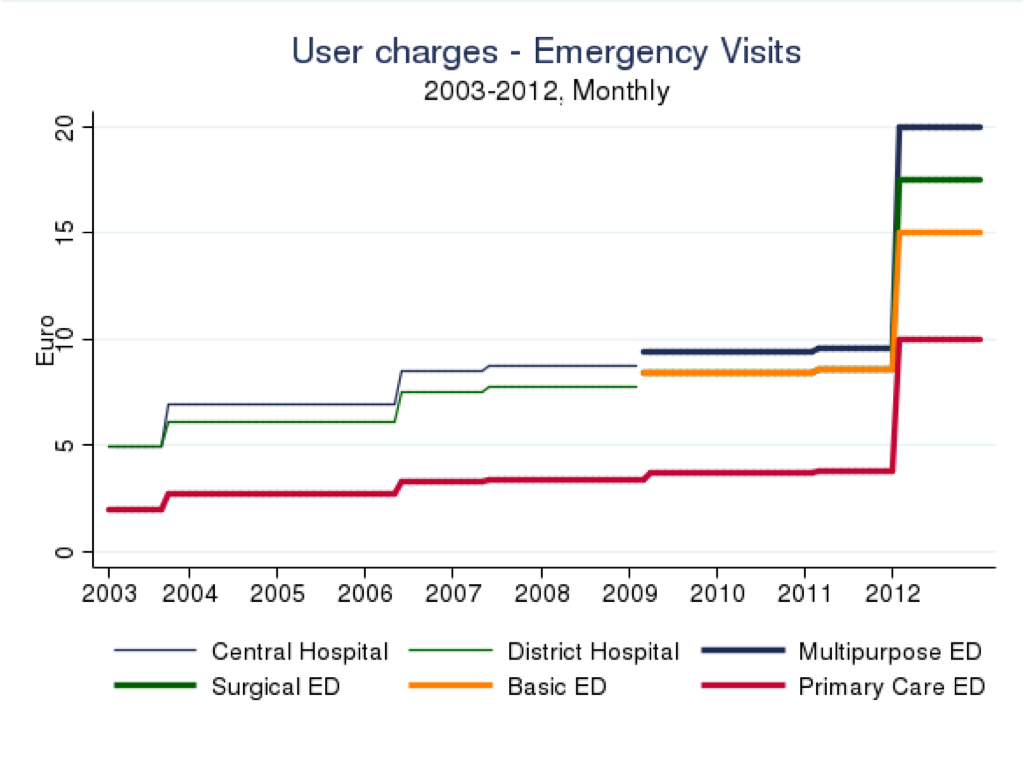}

\input{avg_pcap_don_dist}

\input{avg_pcap_drives_dist}

\input{avg_donations}

\input{avg_drives}

\input{avg_weekendays}

\input{drives_time}

\input{drives_type}

\input{drives_type_d3}

\input{hist_don}

\input{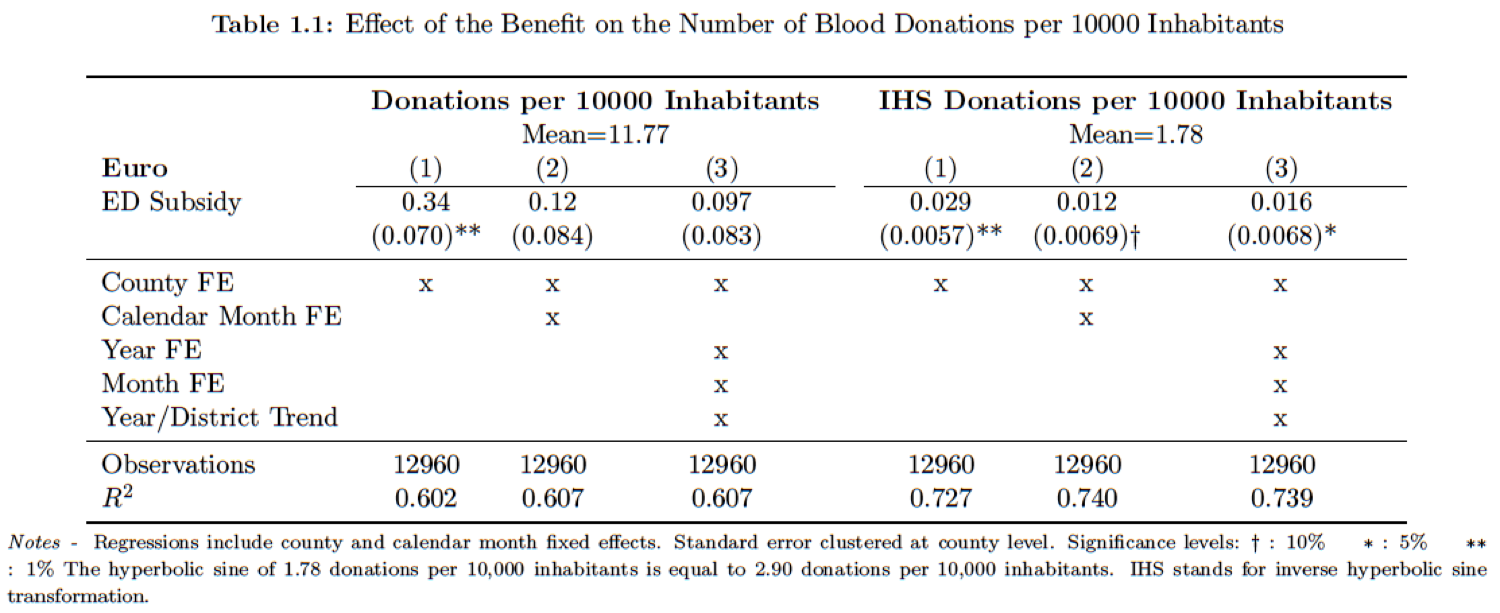}

\input{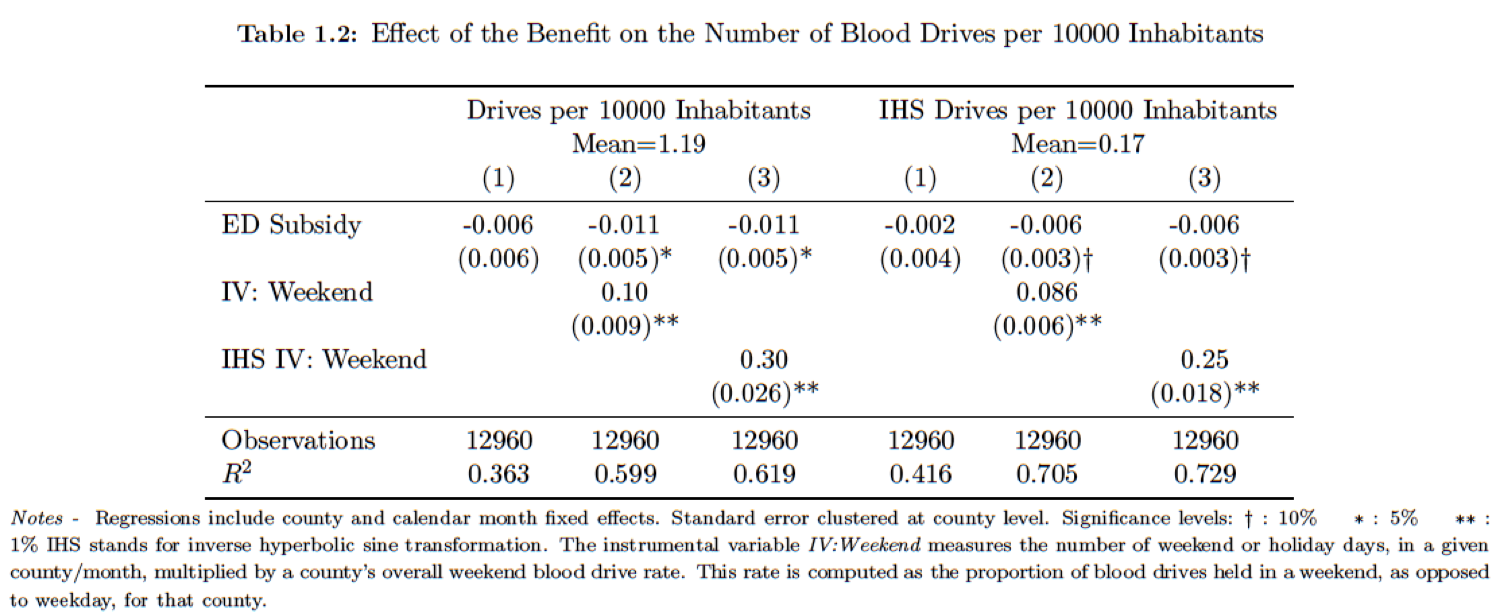}

\input{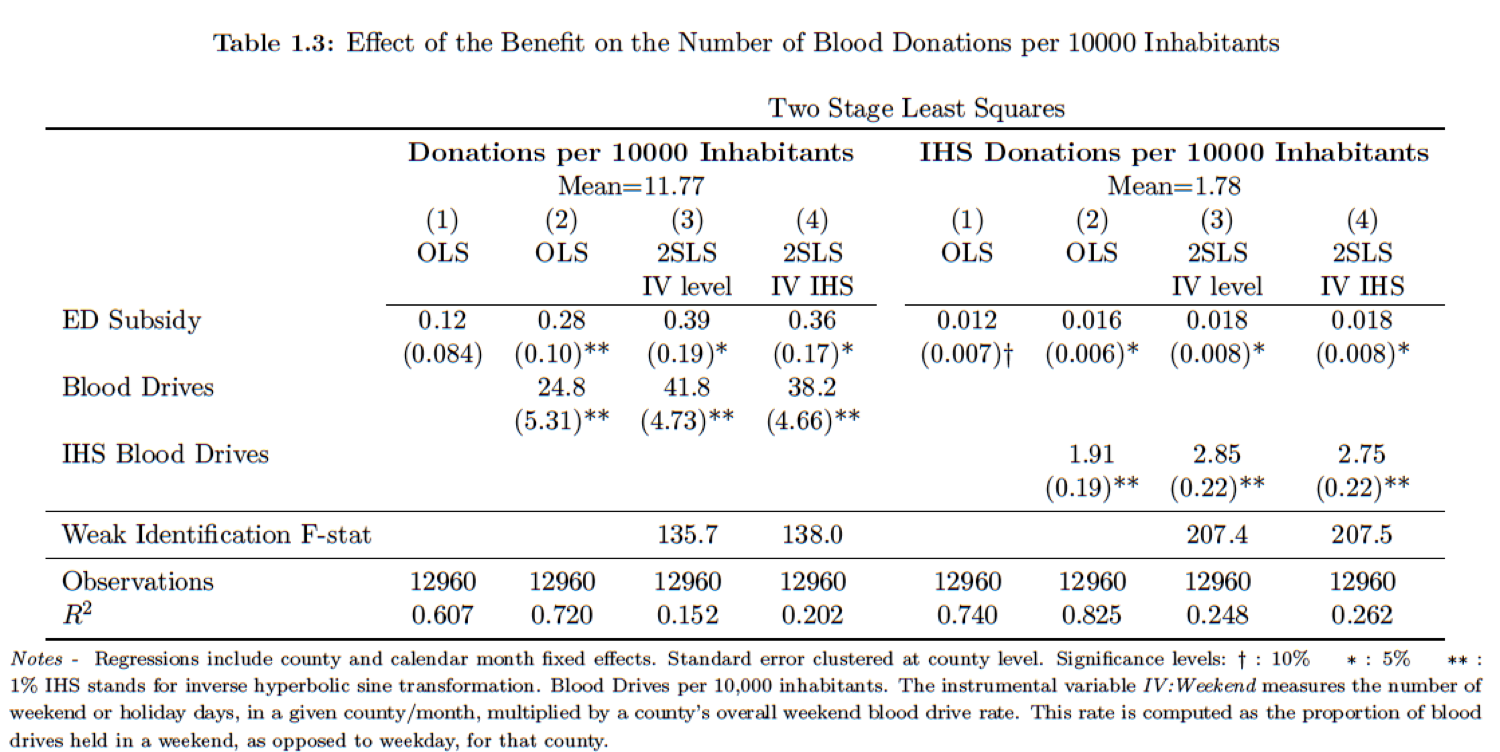}

\input{tab_checks}

\input{tab_sstage_checks}

\input{tab_samples}

\input{tab_sstage_samples}

\input{tab_removal}

\clearpage

\paragraph{Notes about geographical characterization of the data}
There were two sources of geographical information. I was able to identify the town, county and district of the residence of each donor. These data were constructed based on the 4-digit zip code and the name of the town in the zip code in the database. With this information I was able to define the relevant user charge for each donor. Furthermore, I built a similar set of geographical location variables for each blood drive. If its code name was the name of a town, it was assigned the corresponding town-county-district identifiers. Another example would be the name ``Boy Scouts of Azambuja'' (fictional name), where I know that Azambuja is a county in the Lisbon District, which allows to identify geographical location variables. Using this procedure, I identified the location of more than 90\% of the blood drives. As for donations at Blood Donor Centers, I assume they are done at the closest center to the donor's county of residence.


\clearpage
\renewcommand{\bibname}{\thispagestyle{myheadings}References}
\pagebreak

%

\addcontentsline{toc}{chapter}{References} 
\noindent%

\bibliographystyle{apalike}
\bibliography{machado_references}
\thispagestyle{myheadings}

\end{document}

%% file: donations_dc.tex
\let\cleardoublepage\clearpage
	\begin{center}
		\begin{sidewaysfigure}[h]\caption{Blood donations, 2003-2012}\label{donations_dc}
			\centering
				\subfigure[Blood Donations]{\includegraphics[scale=0.22]{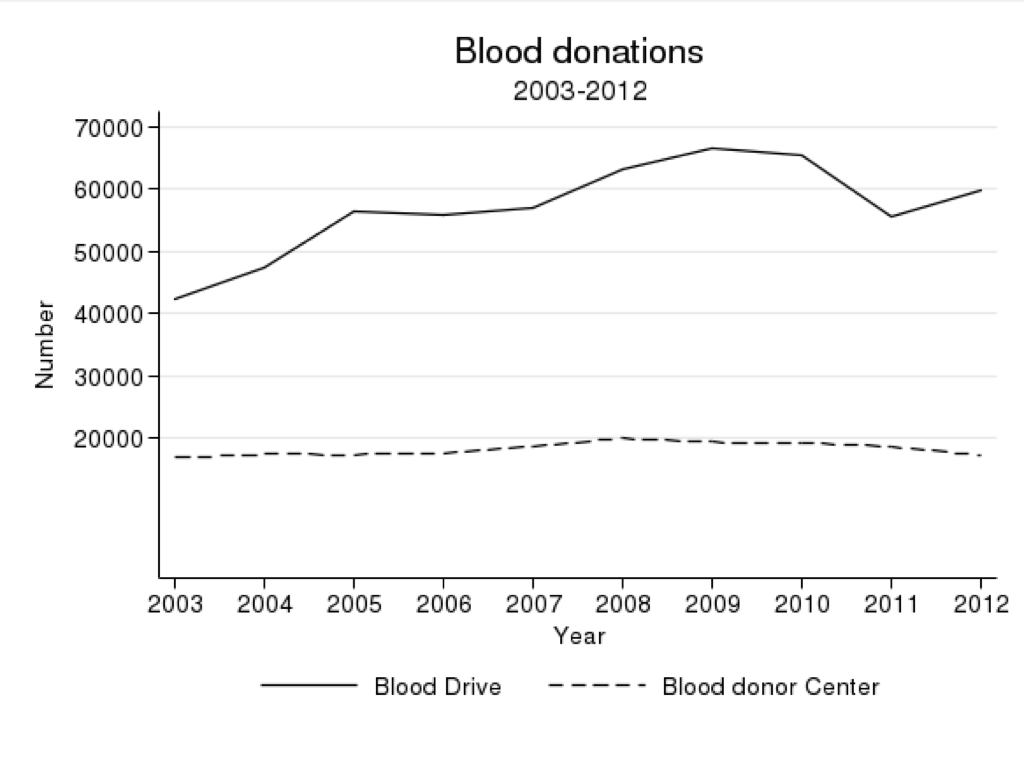}}
			\centering
				\subfigure[Blood Donations - Normalized]{\includegraphics[scale=.22]{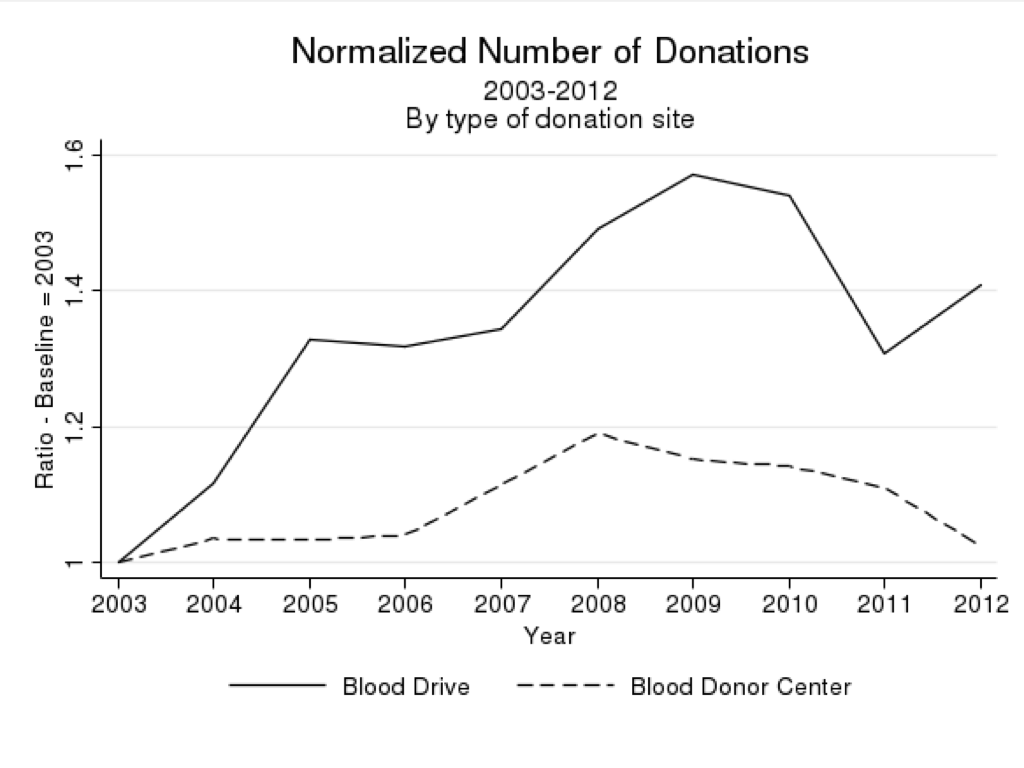}}
			\\
			\begin{mfigsource} 
				Author's compilation based on Portuguese Blood and Transplantation Institute data. Panel B shows the number of donations divided by the number of blood donations in 2003. 
			\end{mfigsource}
		\end{sidewaysfigure}
	\end{center}

%% file: donations_app.tex
\clearpage
	\begin{center}
		\begin{figure}[htb]\caption{Blood donations by triage outcomes, 2003-2012}\label{donations_app}
			\centering
				\includegraphics[scale=0.4]{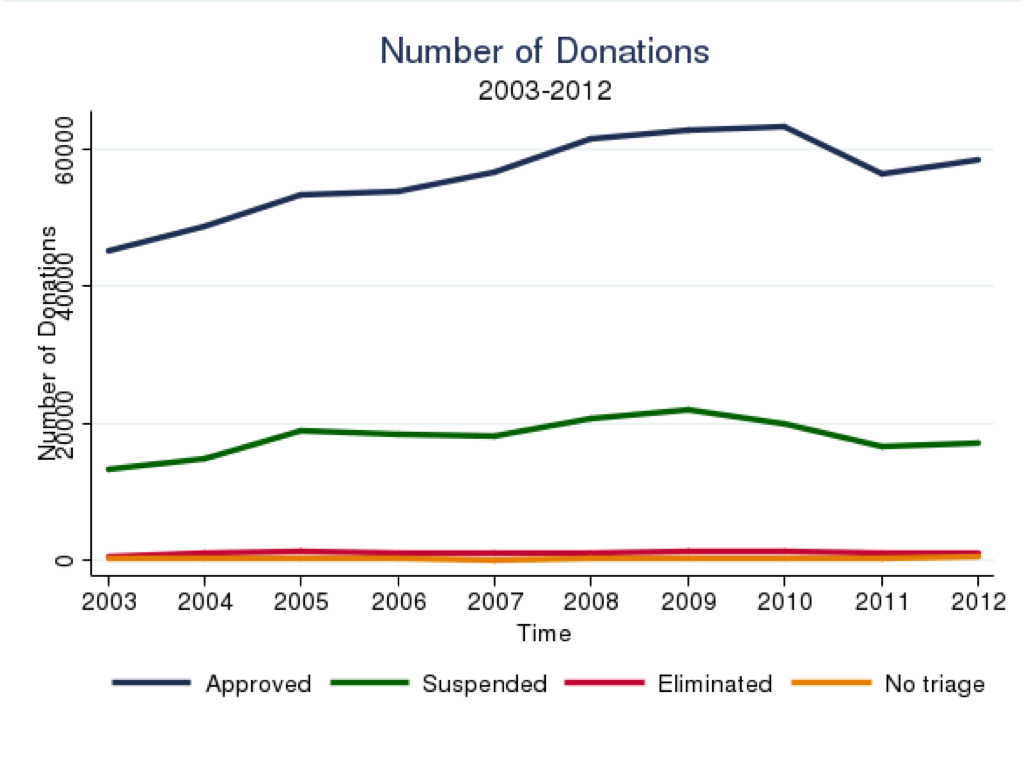}
			\\
			\begin{mfigsource} 
				Author's compilation based on Portuguese Blood and Transplantation Institute data.
			\end{mfigsource}
		\end{figure}
	\end{center}

%% file: drives_dist.tex
\clearpage
\begin{center}
	\begin{figure}[!ht]
	\caption{Blood drives by district, 2003-2012}\label{drives_dist}
	\centering
	\includegraphics[scale=0.40]{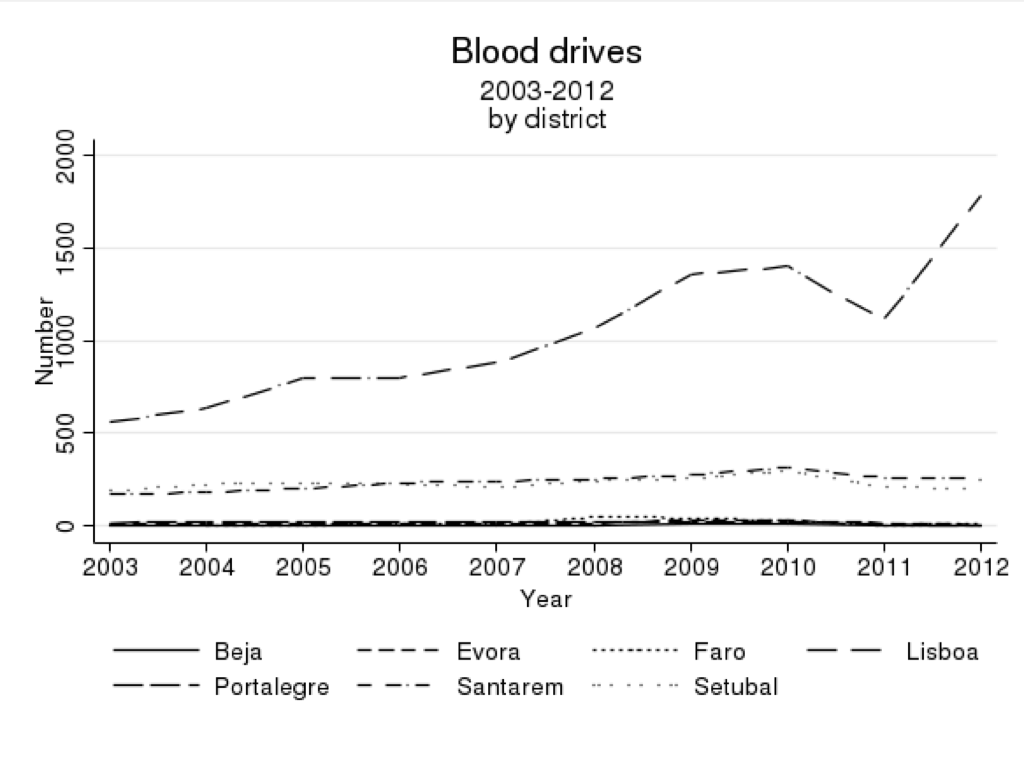} 
	\\
		\begin{mfigsource} 
			Author's compilation based on Portuguese Blood and Transplantation Institute data.
		\end{mfigsource}
	\end{figure}
\end{center}

%% file: usercharges_ed.tex
\let\cleardoublepage\clearpage
	\begin{center}
		\begin{sidewaysfigure}[h]\caption[User fees, 2003-2012]{User fees for visits to the National Health Service, 2003-2012}\label{usercharges_ed}
			\centering
				\subfigure[Emergency Department]{\includegraphics[scale=.22]{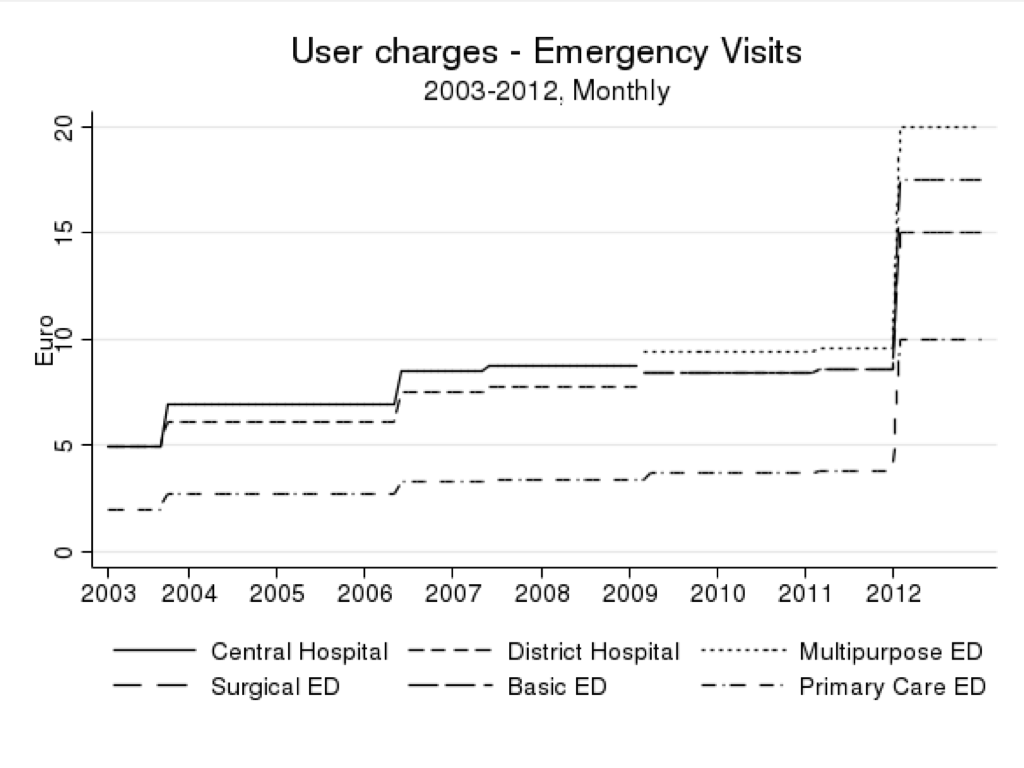}}
			\centering
				\subfigure[Outpatient Appointment]{\includegraphics[scale=.22]{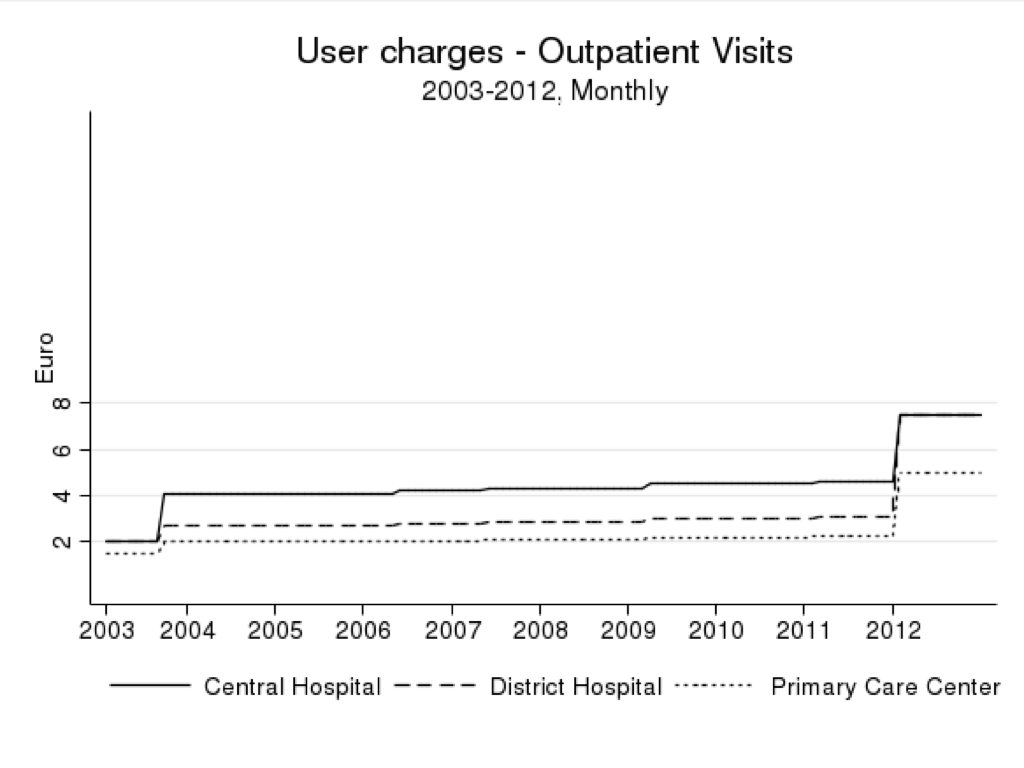}}
			\\
			\begin{mfigsource} 
				Law Decrees regarding changes in user fees, 1993-2012.
			\end{mfigsource}
		\end{sidewaysfigure}
	\end{center}

%% file: avg_pcap_don_dist.tex
\clearpage
\begin{center}
	\begin{figure}[!ht]
	\caption[County/Month blood donations per 10,000 inhabitants, by district]{Average County/Month Blood Donations by 10,000 inhabitants, by district, 2003-2012}\label{avg_pcap_don_dist}
	\centering
	\includegraphics[scale=0.65]{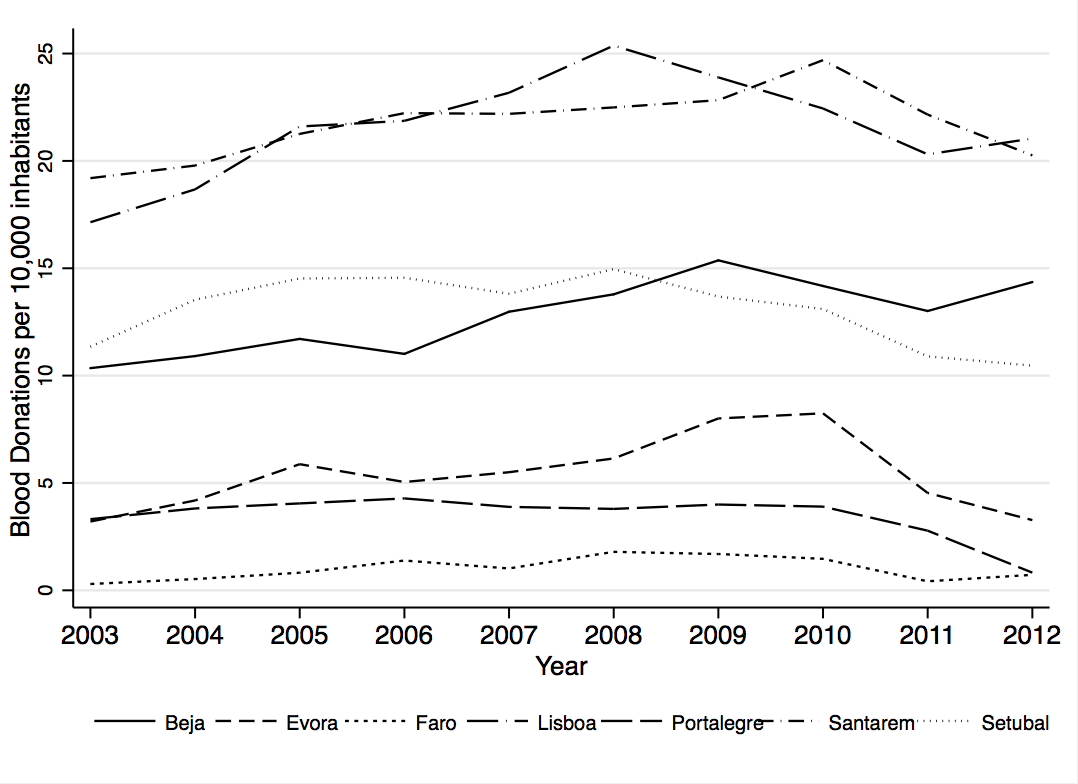} 
	\\
		\begin{mfigsource} 
			Author's compilation based on Portuguese Blood and Transplantation Institute data.
		\end{mfigsource}
	\end{figure}
\end{center}

%% file: avg_pcap_drives_dist.tex
\clearpage
\begin{center}
	\begin{figure}[!ht]
	\caption[County/Month Blood drives per 10,000 inhabitants, by district]{Average County/Month Blood Drives by 10,000 inhabitants, by district, 2003-2012}\label{avg_pcap_drives_dist}
	\centering
	\includegraphics[scale=0.65]{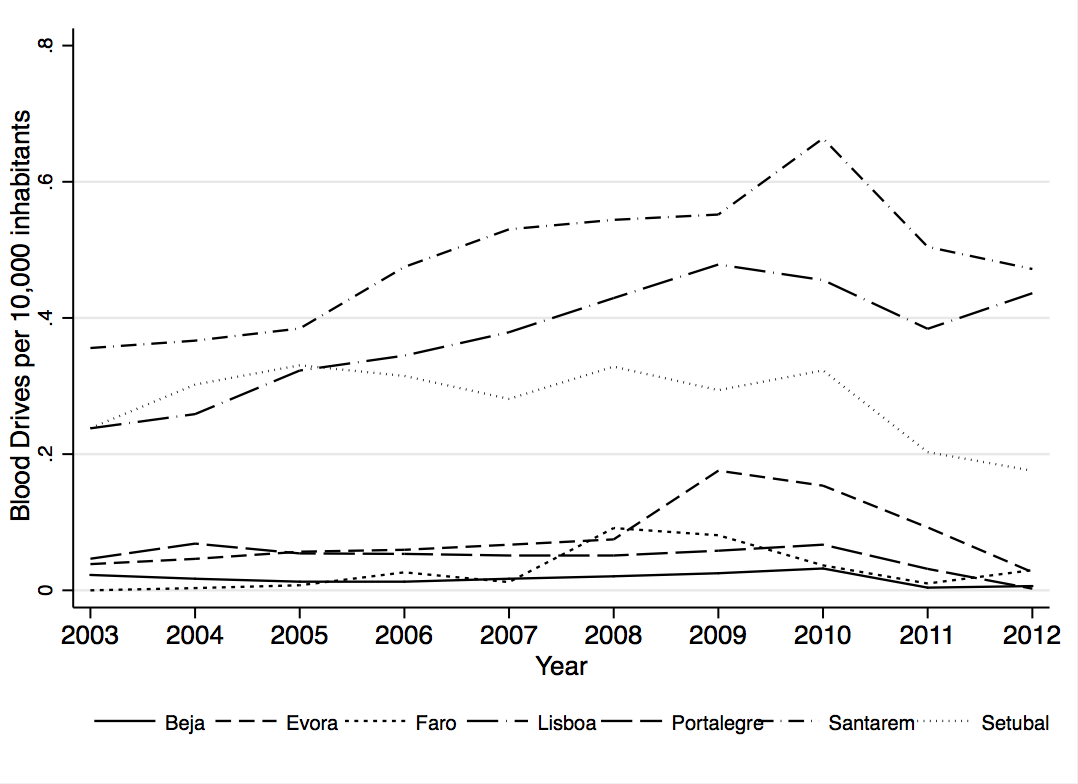} 
	\\
		\begin{mfigsource} 
			Author's compilation based on Portuguese Blood and Transplantation Institute data.
		\end{mfigsource}
	\end{figure}
\end{center}

%% file: avg_donations.tex
\clearpage
	\begin{center}
		\begin{figure}[htb]\caption{User charges and Donations}\label{avg_donations}
			\centering
				\includegraphics[scale=0.45]{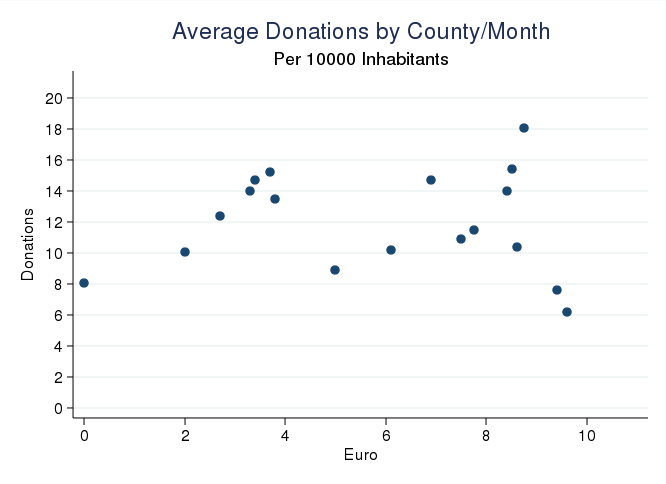}
			\\
			\begin{mfigsource} 
				Author's compilation based on Portuguese Blood and Transplantation Institute data.
			\end{mfigsource}
		\end{figure}
	\end{center}

%% file: avg_drives.tex
\clearpage
	\begin{center}
		\begin{figure}[htb]\caption{User charges and Blood Drives}\label{avg_drives}
			\centering
				\includegraphics[scale=0.45]{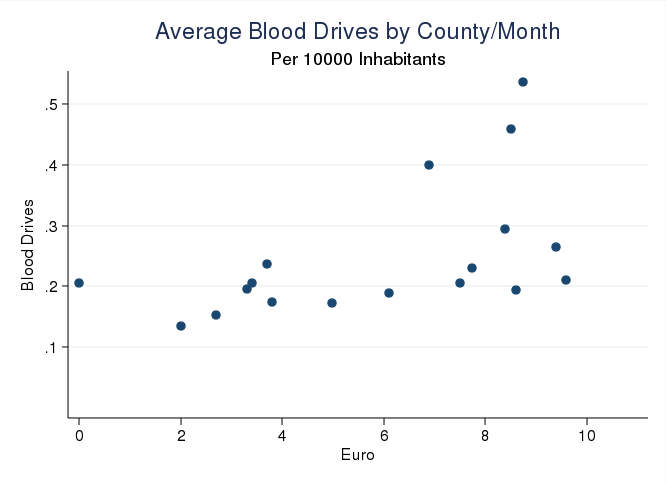}
			\\
			\begin{mfigsource} 
				Author's compilation based on Portuguese Blood and Transplantation Institute data.
			\end{mfigsource}
		\end{figure}
	\end{center}	

%% file: avg_weekendays.tex
\clearpage
	\begin{center}
		\begin{figure}[htb]\caption[Weekend/Holiday Days per Month]{Average Number of Week/Holiday Days per Month, by County}\label{avg_weekendays}
			\centering
				\includegraphics[scale=0.45]{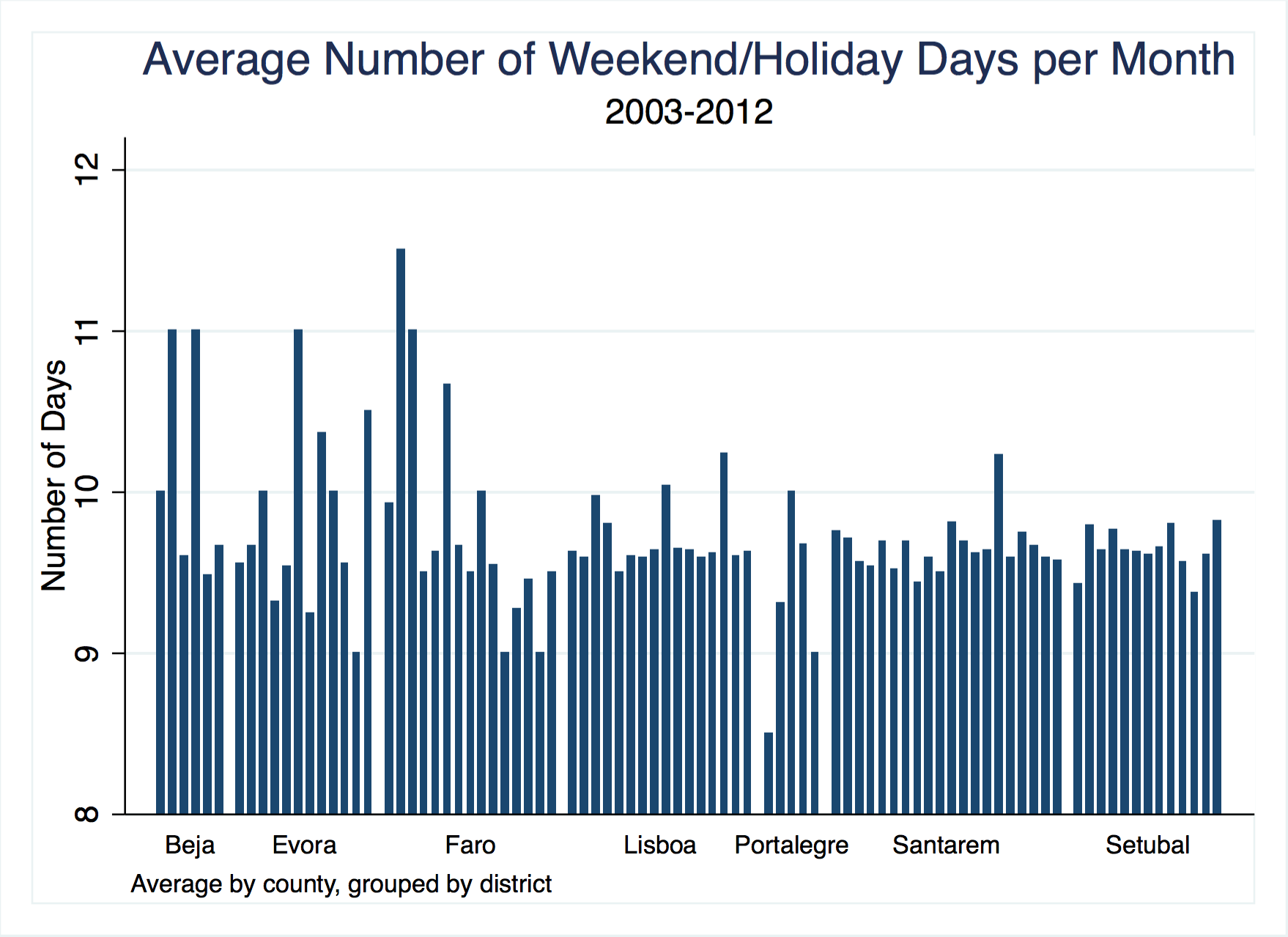}
			\\
			\begin{mfigsource} 
				Author's compilation based on Portuguese Blood and Transplantation Institute data.
			\end{mfigsource}
		\end{figure}
	\end{center}	

%% file: drives_time.tex
\clearpage
	\begin{center}
		\begin{figure}[htb]\caption{Weekend vs Weekday Blood Drives, 2003/2012}\label{drives_time}
			\centering
				\includegraphics[scale=0.45]{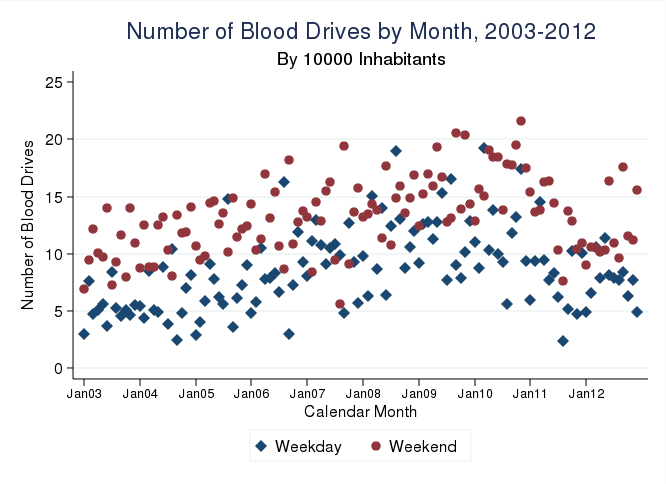}
			\\
			\begin{mfigsource} 
				Author's compilation based on Portuguese Blood and Transplantation Institute data.
			\end{mfigsource}
		\end{figure}
	\end{center}	

%% file: drives_type.tex
\clearpage
	\begin{center}
		\begin{figure}[htb]\caption{Blood Drives - Weekend vs Weekday}\label{drives_type}
			\centering
				\includegraphics[scale=0.60]{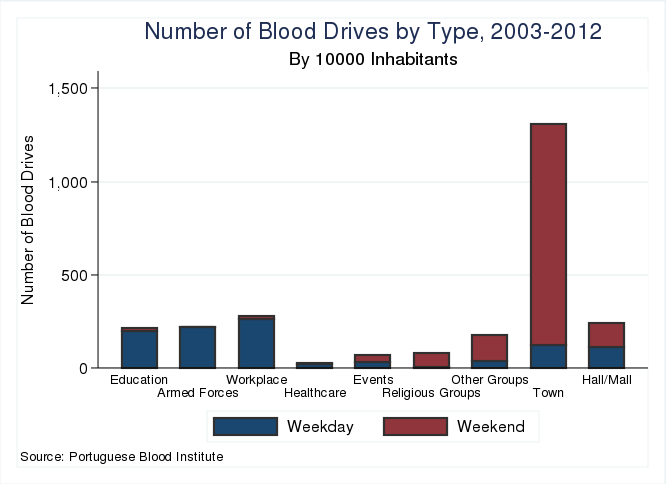}
			\\
			\begin{mfigsource} 
				Author's compilation based on Portuguese Blood and Transplantation Institute data.
			\end{mfigsource}
		\end{figure}
	\end{center}	

%% file: drives_type_d3.tex
\clearpage
	\begin{center}
		\begin{figure}[htb]\caption{Blood Drives - Weekend vs Weekday, Lisbon District, by County}\label{drives_type_d3}
			\centering
				\includegraphics[scale=0.45]{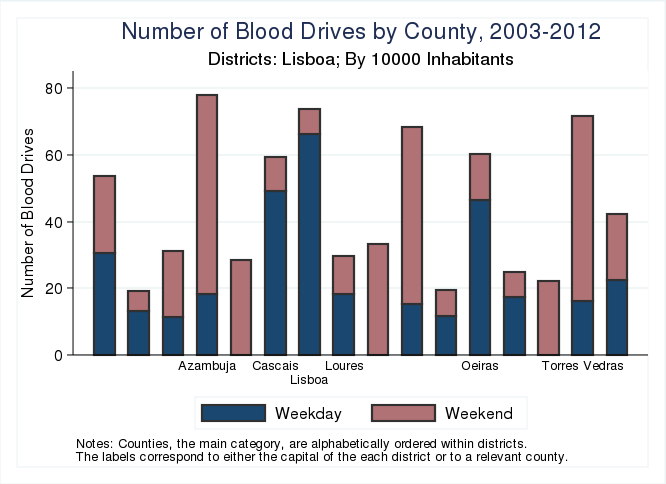}
			\\
			\begin{mfigsource} 
				Author's compilation based on Portuguese Blood and Transplantation Institute data.
			\end{mfigsource}	
		\end{figure}		
	\end{center}

%% file: hist_don.tex
\clearpage
	\begin{center}
		\begin{figure}[htb]\caption{Distribution of Donations per 10,000 inhabitants}\label{hist_don}
			\centering
				\includegraphics[scale=0.45]{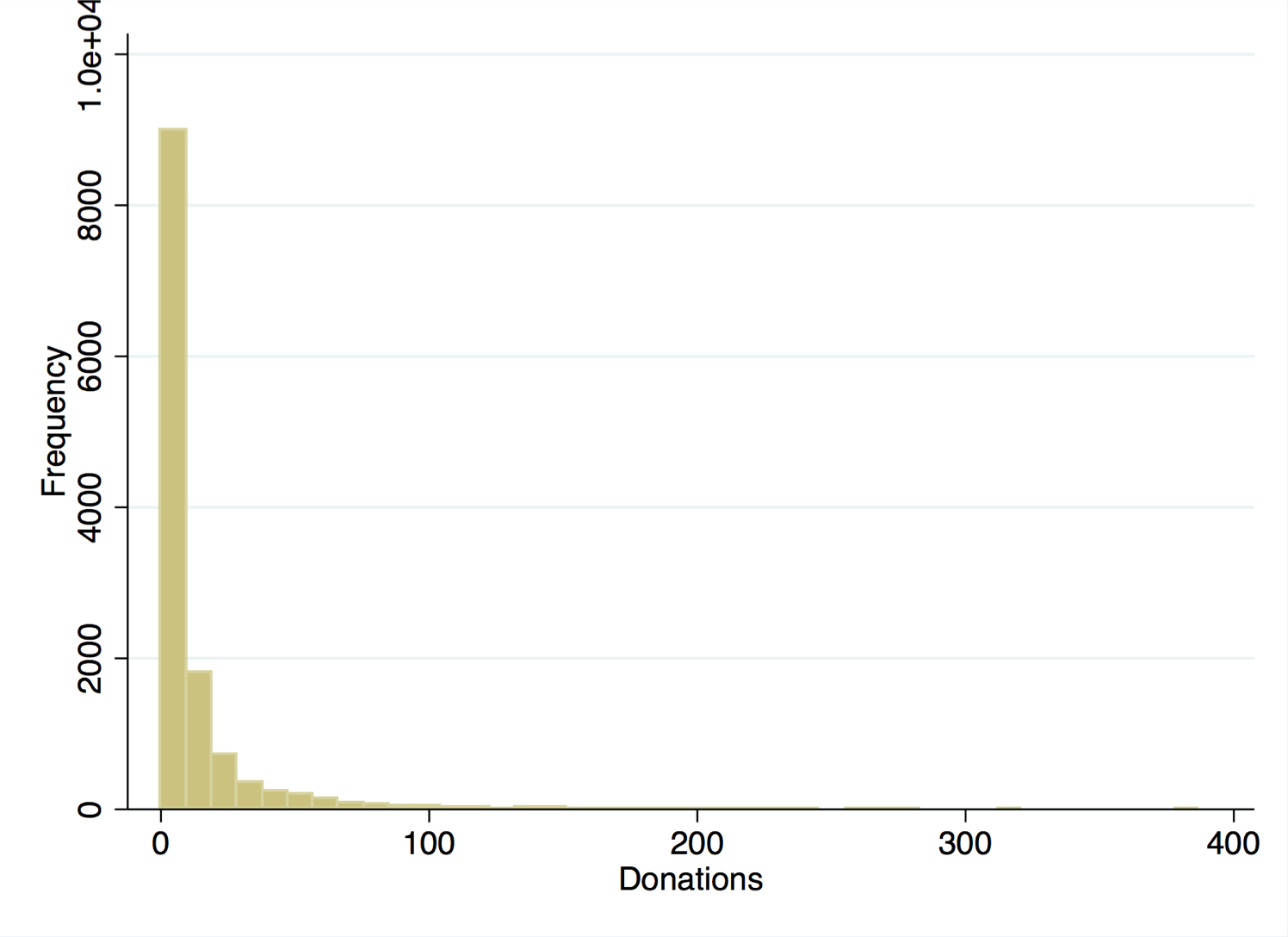}
			\\
			\begin{mfigsource} 
				Author's compilation based on Portuguese Blood and Transplantation Institute data.
			\end{mfigsource}
		\end{figure}
	\end{center}

%% file: tab_all.tex
\clearpage

\begin{sidewaystable}[htbp]\centering
\def\sym#1{\ifmmode^{#1}\else\(^{#1}\)\fi}
\caption[Effect of Benefit on Donations per 10,000 Inhabitants]{Effect of the Benefit on the Number of Blood Donations per 10000 Inhabitants}\label{tab_all}
\bigskip

\begin{tabular}{l*{7}{c}}
\toprule
               &\multicolumn{3}{c}{\tbf{Donations per 10000 Inhabitants} }  & & \multicolumn{3}{c}{\tbf{IHS Donations per 10000 Inhabitants} }\\
              & \multicolumn{3}{c}{Mean=11.77}  & & \multicolumn{3}{c}{Mean=1.78} \\
\tbf{Euro}                    &\multicolumn{1}{c}{(1)}&\multicolumn{1}{c}{(2)}&\multicolumn{1}{c}{(3)}& &\multicolumn{1}{c}{(1)}&\multicolumn{1}{c}{(2)}&\multicolumn{1}{c}{(3)}\\ \cline{2-4} \cline{6-8}

ED Subsidy          &        0.34  &        0.12  &       0.097  & &      0.029  &       0.012  &       0.016  \\
                    &     (0.070)**&     (0.084)  &     (0.083)  &    & (0.0057)**&    (0.0069)\dag&    (0.0068)* \\                    
\midrule
County FE       &  x  &       x     & x   & &       x    &       x   & x    \\
Calendar Month FE  &   &        x   &   & & & x &           \\
Year FE              &             &            &     x     & & & & x      \\
Month FE              &             &            &     x     & & & & x    \\
Year/District  Trend           &             &            &     x     & & & & x      \\
\midrule
Observations        &       12960  &       12960  &       12960  & &       12960  &       12960  &       12960  \\
\(R^{2}\)           &       0.602  &       0.607  &       0.607  &   &    0.727  &       0.740  &       0.739  \\
\bottomrule
\end{tabular}
	\begin{mfignotes} 
			Regressions include county and calendar month fixed effects. Standard error clustered at county level. Significance levels:  $\dag$ : 10\% \hspace{1em} $\ast$ : 5\% \hspace{1em} $\ast\ast$ : 1\%
			The hyperbolic sine of 1.78 donations per 10,000 inhabitants is equal to 2.90 donations per 10,000 inhabitants. IHS stands for inverse hyperbolic sine transformation.  
	\end{mfignotes}

\end{sidewaystable}

%% file: tab_fstage.tex
\clearpage

\begin{sidewaystable}[htbp]\centering
\def\sym#1{\ifmmode^{#1}\else\(^{#1}\)\fi}
\caption[Effect of Benefit on Drives per 10,000 Inhabitants]{Effect of the Benefit on the Number of Blood Drives per 10000 Inhabitants}\label{tab_fstage}

\bigskip

\begin{tabular}{l*{7}{c}}
\toprule
               &\multicolumn{3}{c}{Drives per 10000 Inhabitants} &  &\multicolumn{3}{c}{IHS Drives per 10000 Inhabitants} \\
           & \multicolumn{3}{c}{Mean=1.19}  &&  \multicolumn{3}{c}{Mean=0.17} \\
                  &\multicolumn{1}{c}{(1)}&\multicolumn{1}{c}{(2)} &\multicolumn{1}{c}{(3)}& &\multicolumn{1}{c}{(1)}&\multicolumn{1}{c}{(2)} &\multicolumn{1}{c}{(3)}\\
\midrule
ED Subsidy          &     -0.006  &      -0.011  &      -0.011  &  &   -0.002  &     -0.006  &     -0.006  \\
                    &    (0.006)  &    (0.005)* &    (0.005)* &   & (0.004)  &    (0.003)\dag&    (0.003)\dag\\
IV: Weekend            &              &        0.10  &      &        &              &       0.086  &              \\
                    &              &    (0.009)**&              & &             &    (0.006)**&              \\
IHS IV: Weekend         &              &              &        0.30  &    &          &              &        0.25  \\
                    &              &              &     (0.026)**&              &    &          &     (0.018)**\\
\midrule
Observations        &       12960  &       12960  &       12960  &    &   12960  &       12960  &       12960  \\
\(R^{2}\)           &       0.363  &       0.599  &       0.619  &     &  0.416  &       0.705  &       0.729  \\

\bottomrule
\end{tabular}
	\begin{mfignotes} 
			Regressions include county and calendar month fixed effects. Standard error clustered at county level. Significance levels:  $\dag$ : 10\% \hspace{1em} $\ast$ : 5\% \hspace{1em} $\ast\ast$ : 1\%
			IHS stands for inverse hyperbolic sine transformation.
			The instrumental variable {\em IV:Weekend} measures the number of weekend or holiday days, in a given county/month, multiplied by a county's overall weekend blood drive rate. 
			This rate is computed as the proportion of blood drives held in a weekend, as opposed to weekday, for that county. 
	\end{mfignotes}

\end{sidewaystable}

%% file: tab_sstage.tex
\clearpage

\begin{sidewaystable}[htbp]\centering
\def\sym#1{\ifmmode^{#1}\else\(^{#1}\)\fi}
\caption[Effect of Benefit on Blood Donations per 10000 Inhabitants - Two Stage Least Squares]{Effect of the Benefit on the Number of Blood Donations per 10000 Inhabitants}\label{tab_sstage}

\bigskip

\begin{tabular}{l*{9}{c}}

	&\multicolumn{9}{c}{Two Stage Least Squares}\\   
\toprule
               &\multicolumn{4}{c}{\tbf{Donations per 10000 Inhabitants}} &  &\multicolumn{4}{c}{\tbf{IHS Donations per 10000 Inhabitants}} \\
              & \multicolumn{4}{c}{Mean=11.77}  & & \multicolumn{4}{c}{Mean=1.78} \\
	                &\multicolumn{1}{c}{(1)}&\multicolumn{1}{c}{(2)}&\multicolumn{1}{c}{(3)} &\multicolumn{1}{c}{(4)}& &\multicolumn{1}{c}{(1)}&\multicolumn{1}{c}{(2)}&\multicolumn{1}{c}{(3)} &\multicolumn{1}{c}{(4)}\\
	                &\multicolumn{1}{c}{OLS}&\multicolumn{1}{c}{OLS}&\multicolumn{1}{c}{2SLS} &\multicolumn{1}{c}{2SLS}& &\multicolumn{1}{c}{OLS}&\multicolumn{1}{c}{OLS}&\multicolumn{1}{c}{2SLS} &\multicolumn{1}{c}{2SLS}\\
	                & & &\multicolumn{1}{c}{IV level} &\multicolumn{1}{c}{IV IHS}& & & &\multicolumn{1}{c}{IV level} &\multicolumn{1}{c}{IV IHS}\\ \cline{2-5} \cline{7-10}

ED Subsidy          &        0.12  &        0.28  &        0.39  &        0.36  &    &   0.012  &       0.016  &       0.018  &       0.018  \\
                    &     (0.084)  &      (0.10)**&      (0.19)* &      (0.17)* &  &  (0.007)\dag&    (0.006)* &    (0.008)* &    (0.008)* \\
Blood Drives          &              &        24.8  &        41.8  &        38.2  &              &              &              &              \\
                    &              &      (5.31)**&      (4.73)**&      (4.66)**&              &              &              &              \\
IHS Blood Drives   &              &              &              &              &          &    &        1.91  &        2.85  &        2.75  \\
                    &              &              &              &              &              &    &  (0.19)**&      (0.22)**&      (0.22)**\\
 \midrule
Weak Identification F-stat       &              &              &       135.7  &       138.0  &      &        &              &       207.4  &       207.5  \\
                    \midrule 
Observations       &       12960  &       12960  &       12960  &       12960  &  &     12960  &       12960  &       12960  &       12960  \\
\(R^{2}\)           &       0.607  &       0.720  &       0.152  &       0.202  &    &   0.740  &       0.825  &       0.248  &       0.262  \\
\bottomrule
\end{tabular}
	\begin{mfignotes} 
			Regressions include county and calendar month fixed effects. Standard error clustered at county level. Significance levels:  $\dag$ : 10\% \hspace{1em} $\ast$ : 5\% \hspace{1em} $\ast\ast$ : 1\%
			IHS stands for inverse hyperbolic sine transformation. Blood Drives per 10,000 inhabitants.
			The instrumental variable {\em IV:Weekend} measures the number of weekend or holiday days, in a given county/month, multiplied by a county's overall weekend blood drive rate. 
			This rate is computed as the proportion of blood drives held in a weekend, as opposed to weekday, for that county. 
	\end{mfignotes}
\end{sidewaystable}

%% file: tab_checks.tex
\clearpage

\begin{sidewaystable}[htbp]\centering
\def\sym#1{\ifmmode^{#1}\else\(^{#1}\)\fi}
\caption[Effect of Benefit on Donations per 10,000 Inhabitants - Robustness Checks]{Effect of the Benefit on the Number of Blood Donations per 10000 Inhabitants}\label{tab_checks}
\bigskip

\begin{tabular}{l*{7}{c}}
\toprule
               &\multicolumn{3}{c}{\tbf{Donations per 10000 Inhabitants} }  & & \multicolumn{3}{c}{\tbf{IHS Donations per 10000 Inhabitants} }\\
                    &\multicolumn{1}{c}{All}&\multicolumn{1}{c}{$<2012$}&\multicolumn{1}{c}{Excluding Lisbon} & &\multicolumn{1}{c}{All}&\multicolumn{1}{c}{$<2012$}&\multicolumn{1}{c}{Excluding Lisbon }\\  
\cline{2-4} \cline{6-8}
\tbf{Euro}   &\multicolumn{1}{c}{(1)}&\multicolumn{1}{c}{(2)}&\multicolumn{1}{c}{(3)} & &\multicolumn{1}{c}{(1)}&\multicolumn{1}{c}{(2)} &\multicolumn{1}{c}{(3)}\\

ED Subsidy          &        0.12  &        0.20  &        0.15   &    &   0.012  &       0.011  &       0.017 \\
                    &     (0.084)  &      (0.11)\dag&     (0.091)\dag&   &  (0.0069)\dag&    (0.0087)  &    (0.008)*  \\
\midrule
Mean Dep.Var              &\multicolumn{1}{c}{11.77}&\multicolumn{1}{c}{11.9}&\multicolumn{1}{c}{10.07}& &\multicolumn{1}{c}{1.78}&\multicolumn{1}{c}{$1.79$}&\multicolumn{1}{c}{3.5} \\ 
\midrule
Observations        &       12960  &       11664  &       11040   &  &     12960  &       11664  &       11040   \\
\(R^{2}\)           &       0.607  &       0.601  &       0.611   &    &   0.740  &       0.738  &       0.690   \\
\bottomrule
\end{tabular}
	\begin{mfignotes} 
			Regressions include county and calendar month fixed effects. Standard error clustered at county level. Significance levels:  $\dag$ : 10\% \hspace{1em} $\ast$ : 5\% \hspace{1em} $\ast\ast$ : 1\%
			IHS stands for inverse hyperbolic sine transformation.
	\end{mfignotes}
\end{sidewaystable}

%% file: tab_sstage_checks.tex
\clearpage

\begin{sidewaystable}[htbp]\centering
\def\sym#1{\ifmmode^{#1}\else\(^{#1}\)\fi}
\caption[Effect of Benefit on Blood Donations per 10000 Inhabitants - Two Stage Least Squares - Robustness Checks]{Effect of the Benefit on the Number of Blood Donations per 10000 Inhabitants}\label{tab_sstage_checks}

\bigskip

\begin{tabular}{l*{7}{c}}

	&\multicolumn{7}{c}{Two Stage Least Squares}\\   
\toprule
               &\multicolumn{3}{c}{\tbf{Donations per 10000 Inhabitants}} &  &\multicolumn{3}{c}{\tbf{IHS Donations per 10000 Inhabitants}} \\
              & \multicolumn{3}{c}{Mean=11.77}  & & \multicolumn{3}{c}{Mean=1.78} \\
	                &\multicolumn{1}{c}{(1)}&\multicolumn{1}{c}{(2)}&\multicolumn{1}{c}{(3)}& &\multicolumn{1}{c}{(1)}&\multicolumn{1}{c}{(2)}&\multicolumn{1}{c}{(3)}\\
	                &\multicolumn{1}{c}{All}&\multicolumn{1}{c}{$<2012$}&\multicolumn{1}{c}{Excluding Lisbon}&  &\multicolumn{1}{c}{All}&\multicolumn{1}{c}{$<2012$}&\multicolumn{1}{c}{Excluding Lisbon}\\ \cline{2-4} \cline{6-8}
ED Subsidy          &        0.39  &        0.32  &        0.30  &   &    0.018  &      0.010  &       0.018  \\
                    &      (0.19)* &      (0.20)\dag&      (0.16)\dag&   & (0.008)* &    (0.009)  &    (0.007)* \\
Drives              &        41.8  &        42.3  &        42.2  &             & &              &              \\
                    &      (4.73)**&      (4.85)**&      (5.50)**&              &&              &              \\
IHS Blood Drives   &              &              &              &       & 2.85  &        2.89  &        3.01  \\
                    &              &              &              &   &   (0.22)**&      (0.22)**&      (0.25)**\\
 \midrule
Weak Identification F-stat   &  135.7  &       137.0  &       107.9  &   &   207.4  &       207.1  &       165.1  \\
                    \midrule 
Observations        &       12960  &       11664  &       11040  &    &   12960  &       11664  &       11040  \\
\(R^{2}\)           &       0.152  &       0.148  &       0.133  &   &    0.248  &       0.253  &       0.244  \\

\bottomrule
\end{tabular}
	\begin{mfignotes} 
			Regressions include county and calendar month fixed effects. Standard error clustered at county level. Significance levels:  $\dag$ : 10\% \hspace{1em} $\ast$ : 5\% \hspace{1em} $\ast\ast$ : 1\%
			IHS stands for inverse hyperbolic sine transformation. Blood Drives per 10,000 inhabitants.
			The instrumental variable {\em IV:Weekend} measures the number of weekend or holiday days, in a given county/month, multiplied by a county's overall weekend blood drive rate. 
			This rate is computed as the proportion of blood drives held in a weekend, as opposed to weekday, for that county. 
	\end{mfignotes}
\end{sidewaystable}

%% file: tab_samples.tex
\clearpage

\begin{sidewaystable}[htbp]\centering
\def\sym#1{\ifmmode^{#1}\else\(^{#1}\)\fi}
\caption[Effect of Benefit on Donations per 10,000 Inhabitants - Subgroups]{Effect of the Benefit on the Number of Blood Donations per 10000 Inhabitants}\label{tab_samples}
\bigskip

\begin{tabular}{l*{5}{c}}
\toprule
               &\multicolumn{5}{c}{\tbf{IHS Donations per 10000 Inhabitants} } \\
                 &\multicolumn{1}{c}{(1)}&\multicolumn{1}{c}{(2)}&\multicolumn{1}{c}{(3)}&\multicolumn{1}{c}{(4)}&\multicolumn{1}{c}{(5)}\\
                    &\multicolumn{1}{c}{Donations}&\multicolumn{1}{c}{New Donors}&\multicolumn{1}{c}{Approved}&\multicolumn{1}{c}{Men}&\multicolumn{1}{c}{Women}\\
                    \midrule
 Mean                   &\multicolumn{1}{c}{1.79}&\multicolumn{1}{c}{0.84}&\multicolumn{1}{c}{1.59}&\multicolumn{1}{c}{1.86}&\multicolumn{1}{c}{1.58}\\   
 \midrule           
ED Subsidy          &       0.012  &    -0.00026  &       0.013  &       0.016  &      0.0096  \\
                    &    (0.0069)\dag&    (0.0059)  &    (0.0068)\dag&    (0.0088)\dag&    (0.0057)\dag\\ 
\midrule
Mean  Donations              &\multicolumn{1}{c}{11.77}&\multicolumn{1}{c}{1.84}&\multicolumn{1}{c}{12.70}&\multicolumn{1}{c}{14.19}&\multicolumn{1}{c}{9.5}\\          
\midrule
Observations        &       12960  &       12960  &       12960  &       12960  &       12960  \\
\(R^{2}\)           &       0.740  &       0.532  &       0.734  &       0.721  &       0.712  \\
\bottomrule
\end{tabular}

\bigskip

\begin{tabular}{l*{5}{c}}
\toprule
               &\multicolumn{5}{c}{\tbf{IHS Donations per 10000 Inhabitants} } \\
                 &\multicolumn{1}{c}{(1)}&\multicolumn{1}{c}{(2)}&\multicolumn{1}{c}{(3)}&\multicolumn{1}{c}{(4)}&\multicolumn{1}{c}{(5)}\\
         \tbf{Age Group}          &\multicolumn{1}{c}{$<25$}&\multicolumn{1}{c}{25-34}&\multicolumn{1}{c}{35-44}&\multicolumn{1}{c}{45-54}&\multicolumn{1}{c}{$>55$}\\
                             \midrule
 Mean                   &\multicolumn{1}{c}{1.79}&\multicolumn{1}{c}{0.84}&\multicolumn{1}{c}{1.59}&\multicolumn{1}{c}{1.86}&\multicolumn{1}{c}{1.58}\\   
 \midrule           
ED Subsidy          &     0.00096  &      0.0100  &       0.011  &       0.012  &       0.014  \\
                    &    (0.0048)  &    (0.0057)\dag&    (0.0070)  &    (0.0056)* &    (0.0071)* \\
        \midrule
Mean  Donations              &\multicolumn{1}{c}{1.59}&\multicolumn{1}{c}{2.73}&\multicolumn{1}{c}{2.97}&\multicolumn{1}{c}{2.61}&\multicolumn{1}{c}{1.88}\\   
\midrule
Observations        &       12960  &       12960  &       12960  &       12960  &       12960  \\
\(R^{2}\)           &       0.562  &       0.693  &       0.688  &       0.656  &       0.621  \\
\bottomrule
\end{tabular}
	\begin{mfignotes} 
			Regressions include county and calendar month fixed effects. Standard error clustered at county level. Significance levels:  $\dag$ : 10\% \hspace{1em} $\ast$ : 5\% \hspace{1em} $\ast\ast$ : 1\%
			IHS stands for inverse hyperbolic sine transformation.
	\end{mfignotes}

\end{sidewaystable}

%% file: tab_sstage_samples.tex
\clearpage

\begin{sidewaystable}[htbp]
\centering
\def\sym#1{\ifmmode^{#1}\else\(^{#1}\)\fi}
\caption[Effect of Benefit on Blood Donations per 10000 Inhabitants - Two Stage Least Squares - Subgroups]{Effect of the Benefit on the Number of Blood Donations per 10000 Inhabitants}\label{tab_sstage_samples}

\bigskip
\begin{small}
\begin{tabular}{l*{5}{c}}

	&\multicolumn{5}{c}{Two Stage Least Squares}\\   
\toprule
              &\multicolumn{5}{c}{\tbf{IHS Donations per 10000 Inhabitants} } \\
                  &\multicolumn{1}{c}{(1)}&\multicolumn{1}{c}{(2)}&\multicolumn{1}{c}{(3)}&\multicolumn{1}{c}{(4)}&\multicolumn{1}{c}{(5)}\\
                    &\multicolumn{1}{c}{Donations}&\multicolumn{1}{c}{New Donors}&\multicolumn{1}{c}{Approved}&\multicolumn{1}{c}{Men}&\multicolumn{1}{c}{Women}\\
                    \midrule
ED Subsidy          &       0.018  &      0.004  &       0.018  &       0.022  &       0.015  \\
                    &    (0.008)* &    (0.007)  &    (0.008)* &    (0.009)* &    (0.009)\dag\\
IHS Blood Drives   &        2.85  &        1.85  &        2.79  &        2.90  &        2.90  \\
                    &      (0.22)**&      (0.14)**&      (0.21)**&      (0.22)**&      (0.22)**\\
                    \midrule
Mean  Donations              &\multicolumn{1}{c}{11.77}&\multicolumn{1}{c}{1.84}&\multicolumn{1}{c}{12.70}&\multicolumn{1}{c}{14.19}&\multicolumn{1}{c}{9.5}\\          
Weak Identification F-stat      &       207.4  &       207.4  &       207.4  &       207.4  &       207.4  \\
                    \midrule 
Observations        &       12960  &       12960  &       12960  &     12960  &       12960    \\
\(R^{2}\)           &       0.248  &       0.215  &       0.258  &       0.213  &       0.238  \\
\bottomrule
\end{tabular}

\bigskip

\begin{tabular}{l*{5}{c}}
\toprule
               &\multicolumn{5}{c}{\tbf{IHS Donations per 10000 Inhabitants} } \\
                 &\multicolumn{1}{c}{(1)}&\multicolumn{1}{c}{(2)}&\multicolumn{1}{c}{(3)}&\multicolumn{1}{c}{(4)}&\multicolumn{1}{c}{(5)}\\
         \tbf{Age Group}          &\multicolumn{1}{c}{$<25$}&\multicolumn{1}{c}{25-34}&\multicolumn{1}{c}{35-44}&\multicolumn{1}{c}{45-54}&\multicolumn{1}{c}{$>55$}\\
                             \midrule
ED Subsidy          &      0.004  &       0.014  &       0.016  &       0.017  &       0.019  \\
                    &    (0.005)  &    (0.007)* &    (0.007)* &    (0.008)* &    (0.008)* \\
IHS Blood Drives  &        1.37  &        2.11  &        2.49  &        2.58  &        2.29  \\
                    &      (0.10)**&      (0.15)**&      (0.21)**&      (0.18)**&      (0.17)**\\
          
\midrule
 Mean  Donations              &\multicolumn{1}{c}{1.59}&\multicolumn{1}{c}{2.73}&\multicolumn{1}{c}{2.97}&\multicolumn{1}{c}{2.61}&\multicolumn{1}{c}{1.88}\\   
Weak Identification F-stat      &       207.4  &       207.4  &       207.4  &       207.4  &       207.4  \\
\midrule
Observations        &       12960  &       12960  &       12960  &       12960  &       12960  \\
\(R^{2}\)           &       0.175  &       0.268  &       0.290  &       0.280  &       0.251  \\
\bottomrule
\end{tabular}
\end{small}
	\begin{mfignotes} 
			Regressions include county and calendar month fixed effects. Standard error clustered at county level. Significance levels:  $\dag$ : 10\% \hspace{1em} $\ast$ : 5\% \hspace{1em} $\ast\ast$ : 1\%
			IHS stands for inverse hyperbolic sine transformation. Blood Drives per 10,000 inhabitants.
			The instrumental variable {\em IV:Weekend} measures the number of weekend or holiday days, in a given county/month, multiplied by a county's overall weekend blood drive rate. 
			This rate is computed as the proportion of blood drives held in a weekend, as opposed to weekday, for that county. 
	\end{mfignotes}
\end{sidewaystable}

%% file: tab_removal.tex
\clearpage

\begin{sidewaystable}[htbp]\centering
\def\sym#1{\ifmmode^{#1}\else\(^{#1}\)\fi}
\caption[Effect of User Fees on Donations per 10,000 Inhabitants, 2011/2012]{Effect of User Fees on the Number of Blood Donations per 10000 Inhabitants, 2011/2012}\label{tab_removal}
\bigskip
\begin{footnotesize}
\begin{tabular}{l*{7}{c}}
	\multicolumn{8}{c}{Unconditional Effect}\\   
\toprule
               &\multicolumn{3}{c}{\tbf{Donations per 10000 Inhabitants} }  & & \multicolumn{3}{c}{\tbf{IHS Donations per 10000 Inhabitants} }\\
              & \multicolumn{3}{c}{Mean=10.89}  & & \multicolumn{3}{c}{Mean=1.73} \\
\tbf{Euro}                    &\multicolumn{1}{c}{(1)}&\multicolumn{1}{c}{(2)}&\multicolumn{1}{c}{(3)}& &\multicolumn{1}{c}{(1)}&\multicolumn{1}{c}{(2)}&\multicolumn{1}{c}{(3)}\\ \cline{2-4} \cline{6-8}

ED Subsidy          &        0.110  &        0.110  &        0.180  & &     0.001  &       0.013  &       0.016  \\
                    &     (0.093)  &      (0.14)  &      (0.14)  &   & (0.006)  &     (0.010)  &     (0.010)  \\                                       
                    \midrule
County FE       &  x  &       x     & x   & &       x    &       x   & x    \\
Calendar Month FE  &   &        x   &   & & & x &           \\
Year FE              &             &            &     x     & & & & x      \\
Month FE              &             &            &     x     & & & & x    \\
Year/District  Trend           &             &            &     x     & & & & x      \\
\midrule
Observations        &        2592  &        2592  &        2592  &   &     2592  &        2592  &        2592  \\
\(R^{2}\)           &       0.665  &       0.669  &       0.669  &     &  0.784  &       0.793  &       0.791  \\
\bottomrule
\end{tabular}
\end{footnotesize}

\bigskip
\begin{footnotesize}

\begin{tabular}{l*{9}{c}}

	\multicolumn{10}{c}{Conditional Effect}\\   
\toprule
               &\multicolumn{4}{c}{\tbf{Donations per 10000 Inhabitants}} &  &\multicolumn{4}{c}{\tbf{IHS Donations per 10000 Inhabitants}} \\
              & \multicolumn{4}{c}{Mean=10.89}  & & \multicolumn{4}{c}{Mean=1.73} \\
	                &\multicolumn{1}{c}{(1)}&\multicolumn{1}{c}{(2)}&\multicolumn{1}{c}{(3)} &\multicolumn{1}{c}{(4)}& &\multicolumn{1}{c}{(1)}&\multicolumn{1}{c}{(2)}&\multicolumn{1}{c}{(3)} &\multicolumn{1}{c}{(4)}\\
	& OLS & 2SLS & OLS & 2SLS & & OLS & 2SLS & OLS & 2SLS \\  \cline{2-5} \cline{7-10}
ED Subsidy          &        0.29  &        0.38  &        0.22  &        0.25  &  &     0.022  &       0.028  &       0.019  &       0.021  \\
                    &      (0.17)  &      (0.25)  &      (0.16)  &      (0.19)  &   &  (0.010)* &     (0.014)* &    (0.0097)\dag&     (0.011)\dag\\

Drives              &        25.4  &        37.8  &              &  &            &        1.21  &        2.10  &              &              \\
                    &      (5.68)**&      (4.31)**&              &      &        &      (0.18)**&      (0.22)**&              &              \\

IHS Blood Drives  &              &              &        34.8  &        45.2  &       &       &              &        1.74  &        2.50  \\
                    &              &              &      (6.37)**&      (5.11)**&              &&              &      (0.19)**&      (0.22)**\\
\midrule
Weak Identification F-statistic            &              &        94.0  &              & 155.4 &       &              &        94.0  &              &       155.4  \\
Observations        &        2592  &        2592  &        2592  &        2592  &     &   2592  &        2592  &        2592  &        2592  \\
\(R^{2}\)           &       0.770  &       0.231  &       0.775  &       0.292  &       & 0.847  &       0.122  &       0.856  &       0.246  \\
	
\bottomrule
\end{tabular}
\end{footnotesize}
	\begin{mfignotes} 
			Conditional effect regressions include county and calendar month fixed effects. Standard error clustered at county level. Significance levels:  $\dag$ : 10\% \hspace{1em} $\ast$ : 5\% \hspace{1em} $\ast\ast$ : 1\%
			IHS stands for inverse hyperbolic sine transformation. 1.73 is equivalent to 2.73 Donations per 10000 Inhabitants. Blood Drives per 10,000 inhabitants.
			The instrumental variable {\em IV:Weekend} measures the number of weekend or holiday days, in a given county/month, multiplied by a county's overall weekend blood drive rate. 
			This rate is computed as the proportion of blood drives held in a weekend, as opposed to weekday, for that county. 
	\end{mfignotes}
\end{sidewaystable}